\documentstyle[12pt]{article}

\begin{document}

\title{$U(2)_{L}\times U(2)_{R}$ Chiral Theory of Mesons}
\author{Bing An Li\\
Department of Physics and Astronomy, University of Kentucky\\
Lexington, KY 40506, USA}

\maketitle

\begin{abstract}
A $U(2)_{L}\times U(2)_{R}$ chiral theory of pseudoscalar, vector,
and axial-vector mesons has been proposed.
VMD has been revealed from this theory.
The physical processes of normal parity
and abnormal parity have been studied by using the same lagrangian and
the universality of coupling has been revealed. Two
new mass relations between vector and axial-vector mesons have been
found. Weinberg's first sum rule and new relations about the amplitude
of $a_{1}$ decay are satisfied. KSFR sum rule is satisfied pretty
well. The $\rho$ pole in pion form factor has been achieved. The
theoretical results of $\rho\rightarrow\pi\pi$, $\omega\rightarrow
\pi\pi$, $a_{1}\rightarrow\rho\pi$ and $\pi\gamma$, $\tau\rightarrow
\rho\nu$, $\tau\rightarrow a_{1}\nu$, $\pi^{0}\rightarrow
\gamma\gamma$, $\omega
\rightarrow\pi\gamma$, $\rho\rightarrow\pi\gamma$, $f_{1}
\rightarrow\rho\pi\pi$,
$f_{1}\rightarrow\eta\pi\pi$, $\rho\rightarrow\eta\gamma$, $\omega
\rightarrow\eta\gamma$ are in good agreement with data. Weinberg's
$\pi\pi$ scattering lengths and slopes and $a^{0}_{2}$, $a^{2}_{2}$,
and $b^{1}_{1}$ have been obtained. Especially, the $\rho$ resonance
in the amplitude $T^{1}_{1}$ of $\pi\pi$ scattering has been revealed
from this theory. Two coefficients of chiral perturbation theory
have been determined and they are close to the values used by
chiral perturbation theory.
\end{abstract}

\newpage
\newcommand{\ssp} {\partial \hspace{-.09in}/}
\newcommand{\sspp} {p \hspace{-.07in}/}
\newcommand{\pa} {\partial}
\newcommand{\ssv} {v \hspace{-.09in}/}
\newcommand{\ssD} {D \hspace{-.10in}/}
\newcommand{\ssa} {a \hspace{-.09in}/}
\newcommand{\ssA} {A \hspace{-.09in}/}
\newcommand{\ssW} {W \hspace{-.12in}/}
\newcommand{\da} {\dagger}
\newcommand{\ssun} {\underline{p} \hspace{-.09in}/}
\newcommand{\ssq} {q \hspace{-.09in}/}

\newpage
Chiral symmetry is
one of the most important features revealed from quantum chromodynamics
(QCD).
The chiral perturbation theory(CPT) is successful in describing
the pseudoscalar meson physics[1,2]. Chiral lagrangian has been
used to study physics of vector and axial-vector mesons before
$QCD$[3]. Weinberg's sum rules[4] of $\rho$ and $a_{1}$
mesons, KSRF sum rule[5] of $\rho$ meson
are also among earlier works in this field.
On the other hand, in ref.[6] in terms of large
$N_{c}$ expansion t`Hooft argues that $QCD$ is equivalent to
a meson theory at low energies. In principle, all mesons should
be included in this meson theory. Of course, in chiral perturbation
theory the effects of other mesons have been included in the
coefficients of the lagrangian up to $O(p^{4})$. As a matter of fact,
in ref.[7,8] the authors have found that the vector meson dominates
the structure of the phenomenological chiral lagrangian.
Various effective theory including
$\rho$ and $a_{1}$ mesons has been studied in last decade[9].
Wess-Zumino lagrangian[10] is an important part of the effective
meson theory. Witten[11] and other authors[12,13] have generalized
Wess-Zumino lagrangian
to include vector and axial-vector mesons by the requiring
gauge invariance.
In refs.[12,14] the generalized Wess-Zumino terms have been used to
study meson physics. In this paper a $U(2)_{L}\times U(2)_{R}$
chiral theory of mesons including pseudoscalar, vector and
axial-vector has been studied. The paper is organized as follows.
1) the formalism of the theory; 2) the definitions of the physical
fields; 3) new mass relations between $\rho$ $a_{1}$ and $\omega$
$f_{1}(1285)$; 4) VDM; 5) the decays of $\rho\rightarrow 2\pi$
and $\omega\rightarrow 2\pi$; 6) pion form factor;
7) the decays of $a_{1}\rightarrow\rho\pi$ and $a_{1}\rightarrow
\gamma\pi$; 8) revisit of Weinberg's sum rules; 9) decays of
$\tau\rightarrow\rho\nu$ and $\tau\rightarrow a_{1}\nu$;
9) $\pi\pi$ scattering and determination of the coefficients
of CPT;
10) decays of $\omega\rightarrow\rho\pi$, $\omega\rightarrow\gamma
\pi$, $\rho\rightarrow\gamma\pi$, and $\pi^{0}\rightarrow\gamma
\gamma$; 11)decays of $f_{1}(1285)$; 12) decays of $\rho\rightarrow
\eta\gamma$ and $\omega\rightarrow\eta\gamma$;
13)Large $N_{c}$ expansion; 14)Dynamical chiral symmetry breaking;
15) Derivative expansion;16) Summary of the results.\\

{\large\bf The formalism of $U(2)_{L}\times U(2)_{R}$ chiral
theory of mesons}\\
In this paper only two flavors
are taken into account and we do not need to worry about the
processes forbidden by OZI rule. $\eta'$ meson will not be
discussed. Therefore, $U(1)$ problem is not an issue of this paper.
In flavor space the mesons are coupled to quarks only.
The background field method is a convenient way of deriving
effective lagrangian of mesons.
The ingredients of this effective meson theory are pseudoscalar
mesons(pions
and $u$ and $d$ quark components of $\eta$),
vector mesons($\rho$ and $
\omega$), axial-vector mesons($a_{1}$ and $f_{1}(1285)$), quarks,
leptons, photon, and W bosons.
Using $U(2)_{L}\times U(2)_{R}$ chiral symmetry and the minimum
coupling principle,
the lagrangian has been constructed as
\begin{eqnarray}
{\cal L}=\bar{\psi}(x)(i\ssp+\ssv+e_{0}Q\ssA+\ssa\gamma_{5}
-mu(x))\psi(x)
+{1\over 2}m^{2}_{0}(\rho^{\mu}_{i}\rho_{\mu i}+
\omega^{\mu}\omega_{\mu}+a^{\mu}_{i}a_{\mu i}+f^{\mu}f_{\mu})
\nonumber \\
+\bar{\psi}(x)_{L}\ssW\psi(x)_{L}
+{\cal L}_{EM}+{\cal L}_{W}+{\cal L}_{lepton}
\end{eqnarray}
where \(a_{\mu}=\tau_{i}a^{i}_{\mu}+f_{\mu}\), \(v_{\mu}=\tau_{i}
\rho^{i}_{\mu}+\omega_{\mu}\), $A_{\mu}$ is photon field,
\(Q={\tau_{3}\over 2}+{1\over 6}\) is the electric charge operator
of $u$ and $d$ quarks,
$W^{i}_{\mu}$ is W boson, and \(u=expi\{\gamma_{5}(\tau_{i}\pi_{i}+
\eta)\}\), $m$ is a parameter. In eq.(1) $u$ can be written as
\begin{equation}
u={1\over 2}(1+\gamma_{5})U+{1\over 2}(1+\gamma_{5})U^{\da},
\end{equation}
where \(U=expi(\tau_{i}\pi_{i}+\eta)\).
Mesons are bound states solutions of $QCD$ and in $QCD$ mesons
are not independent degrees of freedom. Therefore, in eq.(1)
there are no kinetic terms for meson fields. The kinetic terms
of meson fields are generated from quark loops.
Using method of path integral to integrating out
the quark fields, the effective lagrangian of mesons(indicated by M)
are obtained
\begin{equation}
exp \{i\int d^{4}x{\cal L}^{M}\}=\int[d\psi][d\bar{\psi}]
exp \{i\int d^{4}x{\cal L}\}.
\end{equation}
The functional integral has been used and the quark fields are
regulated by proper time method[16]. A review of this method
has been given by Ball[15].
This integral can be done in Euclid space(forget photon and W
boson first)
\begin{equation}
{\cal L}^{M}_{E}=log det{\cal D},
\end{equation}
where
\begin{equation}
{\cal D}=\ssp-i\ssv-i\ssa\gamma_{5}+mu.
\end{equation}
The eq.(4) can be written in two parts
\begin{eqnarray}
{\cal L}^{M}_{E}={\cal L}_{RE}+{\cal L}_{IM},\nonumber \\
{\cal L}_{RE}={1\over 2}log det({\cal D}^{\da}{\cal D}),
\;\;\;{\cal L}_{IM}={1\over 2}
log det({\cal D}/{\cal D}^{\da})
\end{eqnarray}
where
\begin{equation}
{\cal D}^{\da}=-\ssp+i\ssv-i\ssa\gamma_{5}+m\hat{u},\;\;\;
\hat{u}=exp(-i)\gamma_{5}(\tau_{i}\pi_{i}+\eta).
\end{equation}
{}From this effective lagrangian it can be seen that the physical
processes with normal party are described
by ${\cal L}_{RE}$ and the ones with abnormal party
are described by ${\cal L}_{IM}$. In terms of Schwenger's proper time
method[16] we have
\begin{equation}
{\cal L}_{RE}={1\over 2}\int d^{4}xTr\int^{\infty}_{0}{d\tau\over
\tau}e^{-{\cal D}^{\da}{\cal D}},
\end{equation}
where the trace is taken in color, flavor, and Lorentz space.
Inserting a complete set of plane wave and subtracting the divergence
at \(\tau=0\), we obtain
\begin{equation}
{\cal L}_{RE}={1\over 2}\frac{1}{\delta^{D}(0)}
\int d^{D}x\frac{d^{D}p}{(2\pi)^{D}}Tr\int^{\infty}
_{0}{d\tau\over \tau}(e^{-\tau{\cal D}'^{\da}{\cal D}'}
-e^{-\Delta_{0}})\delta^{D}(x-y)|_{y\rightarrow x}
\end{equation}
where
\begin{eqnarray}
{\cal D}'=\ssp+i\sspp-i\ssv-i\ssa\gamma_{5}+mu\nonumber \\
{\cal D}'^{\da}=-\ssp-i\sspp+i\ssv-i\ssa\gamma_{5}+m\hat{u},
\nonumber \\
\Delta_{0}=p^{2}+m^{2}.
\end{eqnarray}
In ref.[15], the Seeley-DeWitt coefficients have been used to
evaluate the expansion series of eq.(9). In this paper we use
dimensional regularization. After completing
the integration over $\tau$ the lagrangian ${\cal L}_{RE}$ reads
\begin{eqnarray}
{\cal L}_{RE}={1\over 2}\int d^{D}x\frac
{d^{D}p}{(2\pi)^{D}}\sum^{\infty}_{n=1}{1\over n}\frac{1}{
(p^{2}+m^{2})^{n}}\nonumber \\
Tr\{(\ssp-i\ssv+i\ssa\gamma_{5})(\ssp-i\ssv-i\ssa
\gamma_{5})+2ip\cdot(\partial-iv-ia\gamma_{5})+m\ssD u \},
\end{eqnarray}
where \(\ssD u=\gamma^{\mu}D_{\mu}u\) and
\begin{equation}
D_{\mu}u=\partial_{\mu}u-i[v_{\mu}, u]+i\{a_{\mu}, u\}.
\end{equation}
To the fourth order in covariant derivatives in Minkofsky space the
lagrangian takes following form
\begin{eqnarray}
\lefteqn{{\cal L}_{RE}=\frac{N_{c}}{(4\pi)^{2}}m^{2}{D\over 4}\Gamma
(2-{D\over 2})TrD_{\mu}UD^{\mu}U^{\da}}\nonumber \\
 & &-{1\over 3}\frac{N_{c}}{(4\pi
)^{2}}{D\over 4}\Gamma(2-{D\over 2})\{2\omega_{\mu\nu}\omega^{\mu\nu}
+Tr\rho_{\mu\nu}\rho^{\mu\nu}+2f_{\mu\nu}f^{\mu\nu}+
Tra_{\mu\nu}a^{\mu\nu}\}\nonumber \\
 & &+{i\over 2}\frac{N_{c}}{(4\pi)^{2}}Tr\{D_{\mu}UD_{\nu}U^{\da}+
D_{\mu}U^{\da}D_{\nu}U\}\rho^{\nu\mu}\nonumber \\
 & &+{i\over 2}\frac{N_{c}}{(4\pi)^{2}}Tr\{D_{\mu}U^{\da}D_{\nu}U-D_{\mu}
UD_{\nu}U^{\da}\}a^{\nu\mu} \nonumber \\
 & &+\frac{N_{c}}{6(4\pi)^{2}}TrD_{\mu}D_{\nu}UD^{\mu}D^{\nu}U^{\da}
\nonumber \\
 & &-\frac{N_{c}}{12(4\pi)^{2}}Tr\{
D_{\mu}UD^{\mu}U^{\da}D_{\nu}UD^{\nu}U^{\da}
+D_{\mu}U^{\da}D^{\mu}UD_{\nu}U^{\da}D^{\nu}U
-D_{\mu}UD_{\nu}U^{\da}D^{\mu}UD^{\nu}U^{\da}\} \nonumber \\
 & &+{1\over 2}m^{2}_{0}(\omega_{\mu}\omega_{\mu}+\rho^{i}_{\mu}
\rho^{i\mu}+a^{i}_{\mu}a^{i\mu}+f_{\mu}f^{\mu}),
\end{eqnarray}
where
\begin{eqnarray*}
D_{\mu}U=\partial_{\mu}U-i[\rho_{\mu}, U]+i\{a_{\mu}, U\},\\
D_{\mu}U^{\da}=\partial_{\mu}U^{\da}-i[\rho_{\mu}, U^{\da}]-
i\{a_{\mu}, U^{\da}\},\\
\omega_{\mu\nu}=\partial_{\mu}\omega_{\nu}-\partial_{\nu}\omega_{\mu},
\\
f_{\mu\nu}=\partial_{\mu}f_{\nu}-\partial_{\nu}f_{\mu},\\
\rho_{\mu\nu}=\partial_{\mu}\rho_{\nu}-\partial_{\nu}\rho_{\mu}
-i[\rho_{\mu}, \rho_{\nu}]-i[a_{\mu}, a_{\nu}],\\
a_{\mu\nu}=\partial_{\mu}a_{\nu}-\partial_{\nu}a_{\mu}
-i[a_{\mu}, \rho_{\nu}]-i[\rho_{\mu}, a_{\nu}],\\
D_{\nu}D_{\mu}U=\partial_{\nu}(D_{\mu}U)-i[\rho_{\nu}, D_{\mu}U]
+i\{a_{\nu}, D_{\mu}U\},\\
D_{\nu}D_{\mu}U^{\da}=\partial_{\nu}(D_{\mu}U^{\da})
-i[\rho_{\nu}, D_{\mu}U^{\da}]
-i\{a_{\nu}, D_{\mu}U^{\da}\}.
\end{eqnarray*}

There are correspondence between the schemes of regularization used
in this paper and in ref.[15]. Upon this correspondence and
transforming the formalism of ref.[15] to Minkofsky space,
it can be found that
this formalism(13) is the same with the one presented in ref.[15].

The imaginary lagrangian(6) describes the physical processes with
abnormal party. It should be the generalized Wess-Zumino lagrangian.
We do not use the method of path integral to find ${\cal L}_{IM}$
and we evaluate the terms of ${\cal L}_{IM}$, which can be used
to study physical processes,
in Minkofsky space directly in this paper.

{\large\bf Defining physical meson fields}\\
In eq.(13) there are divergences. The theory studied in this paper
is an effective theory and it is not renormalizable. In order to build
a physical effective meson theory, the introduction
of a cut-off to the theory is necessary and the cut-off will be
determined in this theory.
We define
\begin{eqnarray}
\frac{F^{2}}{16}=\frac{N_{c}}{(4\pi)^{2}}m^{2}\frac{D}{4}
\Gamma(2-{D\over 2}),\\
g^{2}={8\over 3}\frac{N_{c}}{(4\pi)^{2}}{D\over 4}
\Gamma(2-{D\over 2})={1\over 6}{F^{2}\over m^{2}}.
\end{eqnarray}
The relationship between the cut-off and $F^{2}$, g will be explored.
{}From the kinetic terms of meson fields in eq.(13) we can
see that the meson fields in eq.(13) are not physical.
The physical meson fields can be defined in following ways
that make
the corresponding kinetic terms in the standard form.
\begin{eqnarray}
\pi\rightarrow {2\over f_{\pi}}\pi,\;\;\;\eta\rightarrow
{2\over f_{\eta}}\eta, \nonumber \\
\rho\rightarrow {1\over g}\rho,\;\;\;
\omega\rightarrow {1\over g}\omega,
\end{eqnarray}
where $f_{\pi}$ and $f_{\eta}$ are pion and $\eta$ decay constants,
in chiral limit we take \(f_{\pi}=f_{\eta}\).
Use these substitutions the physical masses of $\rho$ and $\omega$
masons are defined as
\begin{equation}
m^{2}_{\rho}=m^{2}_{\omega}={1\over g^{2}}m^{2}_{0}.
\end{equation}
We can also make the same transformation to
$a_{1}$ and $f_{1}$ fields
\begin{equation}
a^{i}_{\mu}\rightarrow {1\over g}a^{i}_{\mu},\;\;\;
f_{\mu}\rightarrow {1\over g}f_{\mu}.
\end{equation}
However, there are other factors for the normalizations of
axial-vector fields. In eq.(13)
there are mixing between $a^{i}_{\mu}$ and $\partial_{\mu}\pi_{i}$
, $f_{\mu}$ and $\partial_{\mu}\eta$.
In chiral limit the mixing
\[\frac{F^{2}}{2g}\partial_{\mu}\pi_{i}a^{\mu}_{i}.\]
comes from the first term of eq.(13). The transformation
\begin{equation}
a^{i}_{\mu}\rightarrow a^{i}_{\mu}-c\partial_{\mu}\pi^{i}
\end{equation}
has been used to erase the mixing.
In chiral limit, c has been determined by cancelling the mixing
term
\begin{equation}
c=\frac{{F^{2}\over 2g}}{m^{2}_{\rho}+{F^{2}\over g^{2}}}.
\end{equation}
There is similar mixing term between $f_{\mu}$ and $\partial_{\mu}
\eta$ and
the transformation
\[f_{\mu}\rightarrow f_{\mu}-c\partial_{\mu}\eta\]
is used to cancel the mixing term. In chiral limit, c of this
formula is the same with eq.(20).
{}From the term
\[\frac{N_{c}}{6(4\pi)^{2}}TrD_{\nu}D_{\mu}UD^{\nu}D^{\mu}U^{\da}\]
of the lagrangian(13), in chiral limit another term related to the
normalization of $a^{i}_{\mu}$ field has been found,
which can be written as
\begin{equation}
\frac{1}{8\pi^{2}g^{2}}
(\partial_{\mu}a^{i}_{\nu}-\partial_{\nu}a^{i}_{\mu})
(\partial^{\mu}a^{i\nu}-\partial^{\nu}a^{i\mu}).
\end{equation}
Combining this term with the kinetic term of $a^{i}_{\mu}$ in
eq.(13), the physical $a^{i}_{\mu}$ field has been defined as
\begin{equation}
a^{i}_{\mu}\rightarrow {1\over g}(1-{1\over 2\pi^{2}g^{2}})
^{-{1\over 2}}a^{i}_{\mu}.
\end{equation}
In the same way, we obtain the physical $f_{\mu}$ field
\begin{equation}
f_{\mu}\rightarrow {1\over g}(1-{1\over 2\pi^{2}g^{2}})
^{-{1\over 2}}f_{\mu}.
\end{equation}
After the transformations(19), in order to make the kinetic
term of pion in the standard form, in the chiral limit
following equation must be satisfied
\begin{equation}
{F^{2}\over 8}(1-{2c\over g})^{2}+{1\over 2}m^{2}_{\rho}c^{2}=
{f^{2}_{\pi}\over 8}.
\end{equation}
The eq.(24) makes the kinetic term of $\eta$ meson field in standard
form. Eqs.(20,24) can be simplified as
\begin{eqnarray}
c=\frac{f^{2}_{\pi}}{2gm^{2}_{\rho}},\\
{F^{2}\over f^{2}_{\pi}}(1-{2c\over g})=1.
\end{eqnarray}

{\large\bf New mass formulas of vector mesons and its chiral
partners}\\
In lagrangian(1) that vector mesons and axial-vector
masons are chiral partners. However, it can be seen from
eqs.(19,22) that vector and
axial-vector meson fields behave differently. Due to eqs.(22,23),
in the couplings of axial-vector fields to others
there is an additional factor $(1-1/2\pi^{2}g^{2})^{-1/2}$.

The physical masses of vector mesons are defined by eq.(17). For
the masses of axial-vector mesons there are three contributors:
the mass term in the lagrangian(1); the
contribution of the first term of the lagrangian(13), which is
${F^{2}\over g^{2}}$; the normalization factor
$(1-\frac{1}{2\pi^{2}g^{2}})^{-1}$. Put all these three factors
together the $a_{1}$ mass has been found to be
\begin{equation}
(1-\frac{1}{2\pi^{2}g^{2}})m^{2}_{a}={F^{2}\over g^{2}}+m^{2}_{\rho}.
\end{equation}
In the same way, we obtain the mass formula of $f_{1}$ meson
\begin{equation}
(1-\frac{1}{2\pi^{2}g^{2}})m^{2}_{f}={F^{2}\over g^{2}}+m^{2}_{\omega}.
\end{equation}
If we ignore the mass difference of $\rho$ and $\omega$ mesons from
these two mass formulas we obtain
\begin{equation}
m_{f}=m_{a}.
\end{equation}
The deviation of this relation from physical values(\(m_{a}=1.26
GeV\) and \(m_{f}=1.285GeV\)) is about $2\%$.

In chiral limit there are three parameters in this theory,
which can be chosen as $g$,
$f_{\pi}$, and $m_{\rho}$. Take $f_{\pi}$ and $m_{\rho}$ as inputs
and choose
\begin{equation}
g=0.35
\end{equation}
to get a better fits. The couplings in
all the physical processes described by ${\cal L}_{RE}$ and
${\cal L}_{IM}$ are fixed by g and c. This is the universality
of coupling in this theory.

{\large\bf Vector meson dominance(VMD)} \\
Vector meson dominance(VMD) has been revealed from this theory.
{}From eq.(1) it can be seen that except the kinetic term of photon,
photon and vector mesons always appear in the combinations
\begin{equation}
{1\over g}\rho^{0}_{\mu}+{1\over 2}eA_{\mu},\;\;\;
{1\over g}\omega_{\mu}+{1\over 6}eA_{\mu}.
\end{equation}
Therefore, the interaction of photon with other fields can be
found from the
interactions of $\rho^{0}$ or $\omega$  with other fields
by using the substitutions
\begin{eqnarray}
\rho^{0}_{\mu}\rightarrow {1\over 2}egA_{\mu},\nonumber \\
\omega_{\mu}\rightarrow {1\over 6}egA_{\mu}.
\end{eqnarray}
Incorporating photon field into lagrangian(13), from the
kinetic terms of $\rho^{0}$ and $\omega$ mesons in lagrangian(13)
we obtain
\begin{eqnarray}
-{1\over 4}\{\partial_{\mu}(\rho^{0}_{\nu}+{1\over 2}e_{0}gA_{\nu})-
\partial_{\nu}(\rho^{0}_{\mu}+{1\over 2}e_{0}gA_{\mu})\}^{2},
\nonumber \\
-{1\over 4}\{\partial_{\mu}(\omega_{\nu}+{1\over 6}e_{0}gA_{\nu})-
\partial_{\nu}(\omega_{\mu}+{1\over 6}e_{0}gA_{\mu})\}^{2}.
\end{eqnarray}
In order to make the kinetic term of photon field in standard
form, it is
needed to redefine the photon field and the charge to be
\begin{equation}
A_{\mu}\rightarrow (1+\frac{5e^{2}_{0}g^{2}}{18})^{-{1\over 2}}
A_{\mu},\;\;\;
e_{0}\rightarrow e(1+\frac{5e^{2}_{0}g^{2}}{18})^{{1\over 2}},
\;\;\;e_{0}A_{\mu}\rightarrow eA_{\mu}.
\end{equation}
{}From eq.(33) the couplings between photon and vector mesons have
been obtained
\begin{eqnarray}
-{1\over 2}{e\over f_{\rho}}F_{\mu\nu}(\partial_{\mu}\rho_{\nu}-
\partial_{\nu}\rho_{\mu}),\nonumber \\
-{1\over 2}{e\over f_{\omega}}F_{\mu\nu}(\partial_{\mu}
\omega_{\nu}-
\partial_{\nu}\omega_{\mu}),
\end{eqnarray}
where
\begin{equation}
{1\over f_{\rho}}={1\over 2}g,\;\;\;
{1\over f_{\omega}}={1\over 6}g.
\end{equation}
The ratio of ${1\over f_{\rho}}$ to ${1\over f_{\omega}}$ is
$1:{1\over 3}$, that is the same with quark model. The comparison
between theoretical and experimental values of $f_{\rho}$ and
$f_{\omega}$ can be found in Table II.
The photon-vector meson couplings shown
by eqs.(35) are just the ones proposed in ref.[18].
On the other hand, there are interactions between $\rho$ and
$\omega$ mesons with other masons, in general,
these interactions can be written as
\[\rho^{i\mu}j^{i}_{\mu}+\omega^{\mu}j^{\omega}_{\mu}.\]
Therefore,
besides the direct coupling of photon and $\rho$ meson(35)
another type of interaction between photon and other mesons can be
found by the substitution(32)
\[{e\over f_{\rho}}A^{\mu}j^{0}_{\mu}+{e\over f_{\omega}}A_{\mu}
j^{\omega}_{\mu}.\]
The complete expression of the interaction between isovector photon
and mesons is
\begin{equation}
{e\over f_{\rho}}\{-{1\over 2}F^{\mu\nu}(\partial_{\mu}\rho^{0}_
{\nu}-\partial_{\nu}\rho^{0}_{\mu})+A^{\mu}j^{0}_{\mu}\}.
\end{equation}
This is the exact expression of VMD proposed by Sakurai[19]. In the
same way the isoscalar VMD has been found
\begin{equation}
{e\over f_{\omega}}\{-{1\over 2}F^{\mu\nu}(\partial_{\mu}\omega_
{\nu}-\partial_{\nu}\omega_{\mu})+A^{\mu}j^{\omega}_{\mu}\}.
\end{equation}

The cause obtaing the explicit expression of VMD in this theory
can be manifested in another way. It is well known that in $QCD$
the electric current takes the form of
\[\bar{\psi}Q\gamma_{\mu}\psi={1\over 2}\bar{\psi}\tau_{i}
\gamma_{\mu}\psi+{1\over 6}\bar{\psi}\gamma_{\mu}\psi\]
which is in the lagrangian(1). The electric current $\bar{\psi}
Q\gamma_{\mu}\psi$ can be bosonized in this theory.
In ref.[20] we have developed a method to find the effective currents
in the case that only pseudoscalar fields are taken as background
fields. In this paper this method has been generalized to include
vector and axial-vector mesons. From the
lagrangian(1) the equation satisfied by the quark propagator
has been obtained
\begin{equation}
\{i\ssp+{1\over g}\ssv(x)+{1\over g'_{a}}\ssa(x)\gamma_{5}-mu(x)\}
s_{F}(x,y)=\delta
^{4}(x-y),
\end{equation}
where \(g'_{a}=g(1-{1\over 2\pi^{2}g^{2}})^{{1\over 2}}\).
In momentum picture there is
\begin{equation}
s_{F}(x,y)={1\over (2\pi)^{4}}\int d^{4}ye^{-ip(x-y)}s_{F}(x,p).
\end{equation}
The eq.(39) becomes
\begin{equation}
\{i\ssp+\sspp+{1\over g}\ssv(x)+{1\over g'_{a}}\ssa(x)\gamma_{5}
-mu(x)\}s_{F}(x,p)=1.
\end{equation}
Eq.(41) has been solved
\begin{equation}
s_{F}(x,p)=s^{0}_{F}\sum_{n=0}^{\infty}(-)^{n}(
\{i\ssp+{1\over g}\ssv(x)+{1\over g'_{a}}\ssa(x)\gamma_{5}\}
s^{0}_{F})^{n}
\end{equation}
where
\begin{equation}
s^{0}_{F}=-\frac{\sspp-m\hat{u}}{p^{2}-m^{2}}.
\end{equation}
In terms of eqs.(40,42) the bosonization of quark electric current
has been done in following way
\begin{eqnarray}
<\bar{\psi}(x)Q\gamma_{\mu}\psi(x)>=
{1\over 2}<\bar{\psi}(x)\tau_{3}\gamma_{\mu}\psi(x)>+
{1\over 6}<\bar{\psi}(x)\gamma_{\mu}\psi(x)>,\nonumber \\
<\bar{\psi}(x)\tau_{3}\gamma_{\mu}\psi(x)>=-iTr\tau_{3}
\gamma_{\mu}s_{F}(x,x),\;\;\;
<\bar{\psi}(x)\gamma_{\mu}\psi(x)>=-iTr\gamma_{\mu}s_{F}(x,x).
\end{eqnarray}
The leading terms of eq.(44) have been found at \(n=3\)
and the effective currents take following forms
\begin{equation}
<\bar{\psi}\tau_{3}\gamma_{\mu}\psi>=
g\partial^{2}\rho^{0}_{\mu}+gj^{0}_{\mu},\;\;\;
<\bar{\psi}\gamma_{\mu}\psi>=
g\partial^{2}\omega_{\mu}+gj^{\omega}_{\mu}.
\end{equation}
In obtaining the first term of eqs.(45), eq.(15)
has been used. In eq.(45)
$j^{0}_{\mu}$ and $j^{\omega}_{\mu}$ have been defined as
the rest parts of the
currents and $j^{\omega}_{\mu}$ will be evaluated explicitly below.
{}From eq.(1) it can be seen that $j^{0}_{\mu}$ couples to $\rho^{0}
_{\mu}$ and $j^{\omega}_{\mu}$ couples to $\omega$. Therefore,
these two currents are the currents mentioned in eqs.(37,38).
After getting rid of
total derivative terms we have
\begin{eqnarray}
eA^{\mu}<\bar{\psi}Q\gamma_{\mu}\psi>={1\over f_{\rho}}
\{-{1\over 2}F^{\mu\nu}(\partial_{\mu}\rho_{\nu}-\partial_{\nu}
\rho_{\mu})+A^{\mu}j^{0}_{\mu}\}\nonumber \\
+{1\over f_{\omega}}\{-{1\over 2}F^{\mu\nu}(
\partial_{\mu}\omega_{\nu}-\partial_{\nu}\omega_{\mu})+
A^{\mu}j^{\omega}_{\mu}\}).
\end{eqnarray}
This the same with eqs.(37,38).

{\large\bf The decays of $\rho\rightarrow \pi\pi$, $\omega
\rightarrow \pi\pi$, and KSFR sum rule}\\
In this theory pion is associated with $\gamma_{5}$(see eq.(1)) and
only even number of $\gamma_{5}$ involved in
$\rho\pi\pi$ vertex. Therefore, the
vertex of $\rho\pi\pi$ can be found from eq.(13), which is
\begin{equation}
{\cal L}_{\rho\pi\pi}={2\over g}\epsilon_{ijk}\rho^{\mu}_{i}\pi_{j}
\partial_{\mu}\pi_{k}+{2\over \pi^{2}gf^{2}_{\pi}}[4\pi^{2}c^{2}
-(1-{2c\over g})^{2}]\epsilon_{ijk}\rho^{\mu}_{i}\partial_{\nu}
\pi_{j}\partial_{\mu\nu}\pi_{k}.
\end{equation}
In deriving eq.(47), eq.(26) has been used.
For the decay of $\rho\rightarrow\pi\pi$ eq.(47) becomes the
\begin{eqnarray}
{\cal L}_{\rho\pi\pi}=f_{\rho\pi\pi}\epsilon_{ijk}\rho^{\mu}_{i}
\pi_{j}\partial_{\mu}\pi_{k},\nonumber \\
f_{\rho\pi\pi}={2\over g}\{1+\frac{m^{2}_{\rho}}{2\pi^{2}f^{2}_{\pi}}
[(1-{2c\over g})^{2}-4\pi^{2}c^{2}]\}.
\end{eqnarray}
The choice of \(g=0.35\) makes
\begin{equation}
f_{\rho\pi\pi}={2\over g}.
\end{equation}
Using eq.(48) we obtain
\begin{equation}
\Gamma(\rho\rightarrow\pi\pi)=\frac{f^{2}_{\rho\pi\pi}}
{48\pi}m_{\rho}
(1-\frac{4m^{2}_{\pi}}{m^{2}_{\rho}})^{{3\over 2}}=135 MeV.
\end{equation}
The experimental value is 151MeV. The deviation is about $10\%$.
The KSFR sum rule[5]
\begin{equation}
g_{\rho\gamma}={1\over 2}f_{\rho\pi\pi}f_{\pi}
\end{equation}
is the result of current algebra and PCAC. From eq.(35) we have
\begin{equation}
g_{\rho\gamma}={1\over 2}gm^{2}_{\rho}.
\end{equation}
Substituting eqs.(49,52) into the KSFR sum rule we obtain
\begin{equation}
g^{2}=2\frac{f^{2}_{\pi}}{m^{2}_{\rho}},\;\;\; g=0.342.
\end{equation}
Comparing the value of g with (30)
it can be seen that KSFR sum rule is satisfied pretty well.

In chiral limit the mixing between $\omega$ and $\rho$ is caused by
electromagnetic interaction. The lagrangian of this mixing is
\begin{equation}
{\cal L}_{i}=-{1\over 4}egF^{\mu\nu}(\partial_{\mu}\rho_{\nu}-
\partial_{\nu}\rho_{\mu})
-{1\over 12}egF^{\mu\nu}(\partial_{\mu}\omega_{\nu}-
\partial_{\nu}\omega_{\mu}).
\end{equation}
The mixing angle has been found from eq.(54)
\begin{eqnarray}
sin2\theta=\frac{{2\pi\over 3}\alpha g^{2}\bar{m}^{2}_{\rho}}
{m^{2}_{\omega}-m^{2}_{\rho}},\nonumber \\
\theta=1.74^{0},
\end{eqnarray}
where \(\bar{m}^{2}_{\rho}={1\over 2}(m^{2}_{\rho}+m^{2}_{\omega})\).
The decay width is
\begin{equation}
\Gamma(\omega\rightarrow\pi\pi)=sin^{2}\theta\Gamma(\rho\rightarrow
\pi\pi)\frac{p^{*3}}{p^{3}}{m^{2}_{\rho}\over m^{2}_{\omega}}
=0.136MeV,
\end{equation}
where $p^{*}$ is the momentum of pion when the mass of $\rho$ is
$m_{\omega}$ and $p$ is the momentum of pion when the mass of $\rho$
is really $m_{\rho}$. The experimental value is
$0.186(1\pm 0.15)$MeV.

{\large\bf Pion form factor}\\
According to VMD(eq.(37)), the vertex of $\pi\pi\gamma$ consists of
two parts: direct coupling
and the indirect coupling through a $\rho$ meson.
The lagrangian
of direct coupling can be found either from the eq.(13) or by
substituting $\rho^{0}\rightarrow{e\over f_{\rho}}A$ in eq.(48)
\begin{equation}
{\cal L}_{\pi\pi\gamma}=e\frac{f_{\rho\pi\pi}}{f_{\rho}}
\epsilon_{3jk}A^{\mu}\pi_{j}\partial_{\mu}\pi_{k}.
\end{equation}
Due to the coupling $-{1\over 2}{e\over f_{\rho}}F_{\mu\nu}
(\partial^{\mu}\rho^{\nu}-\partial^{\nu}\rho^{\mu})$ the indirect
coupling of $\pi\pi\gamma$ is proportional to $q^{2}$(q is the
momentum of photon), therefore the charge normalization of
$\pi^{+}$ is satisfied
by \(\frac{f_{\rho\pi\pi}}{f_{\rho}}=1\).
{}From the lagrangian
\begin{equation}
{\cal L}=-{1\over 2}{e\over f_{\rho}}\{F^{\mu\nu}(\partial_{\mu}\rho
_{\nu}-\partial_{\nu}\rho_{\mu})\}+{\cal L}_{\pi\pi\gamma}+{\cal L}
_{\rho\pi\pi}.
\end{equation}
The pion form factor has been obtained
\begin{equation}
F_{\pi}(q^{2})=\frac{1}{1-{q^{2}\over m^{2}_{\rho}}}.
\end{equation}
VMD results $\rho$ pole in pion form factor[21]. The radius of pion is
\begin{equation}
\sqrt{<r^{2}>_{\pi}}=0.63fm.
\end{equation}
The experimental value is $0.663\pm 0.023$fm[22].

{\large\bf Decays of $a_{1}\rightarrow\rho\pi$ and $\pi\gamma$}\\
The decay $a_{1}\rightarrow\rho\pi$ is a process with normal parity.
In chiral limit this vertex has been found from eq.(13)
\begin{eqnarray}
\lefteqn{{\cal L}_{a_{1}\rightarrow\rho\pi}=
\epsilon_{ijk}\{Aa^{\mu}_{i}\rho_{j\mu}\pi_{k}
+Ba^{\mu}_{i}\rho^{\nu}_{j}\partial_{\mu\nu}\pi_{k}\},}\nonumber \\
&&A={2\over f_{\pi}}(1-{1\over 2\pi^{2}g^{2}})^{-{1\over 2}}
\{{F^{2}\over g^{2}}+\frac{m^{2}_{a}}{2\pi^{2}g^{2}}
-{2c\over g}(p_{\pi}\cdot p_{\rho}+p_{\pi}\cdot p_{a})-\frac{3}
{2\pi^{2}g^{2}}(1-{2c\over g})p_{\pi}\cdot p_{\rho}\}\nonumber \\
&&={2\over f_{\pi}}(1-{1\over 2\pi^{2}g^{2}})^{-{1\over 2}}
\{{F^{2}\over g^{2}}+\frac{m^{2}_{a}}
{2\pi^{2}g^{2}}-[{2c\over g}+{3\over 4\pi^{2}g^{2}}(1-{2c\over g})]
(m^{2}_{a}-m^{2}_{\rho})\}\nonumber \\
&&={2\over f_{\pi}}(1-{1\over 2\pi^{2}g^{2}})^{-{1\over 2}}
(m^{2}_{a}-m^{2}_{\rho})(1-{2c\over g})(1-{3\over
4\pi^{2}g^{2}}),\nonumber \\
&&B=-{2\over f_{\pi}}(1-{1\over 2\pi^{2}g^{2}})^{-{1\over 2}}
{1\over 2\pi^{2}g^{2}}(1-{2c\over g}).
\end{eqnarray}
The three expressions of A in eq.(61) have different uses in this
paper and in obtaining the last two expressions of A eq.(27) has
been used.
The width of the decay has been calculated and is
326MeV which is comparable with data[17].
{}From eq.(61) it can be seen that there are s-wave and d-wave in this
decay. The ratio[23] of these two waves obtained in this theory is
\begin{equation}
{s\over d}=-{1\over 3}p^{2}_{\pi}\frac{\frac{A}{m_{\rho}(m_{\rho}
+E_{\rho})}-B{m_{a}\over m_{\rho}}}{A(1+{1\over 3}\frac{p^{2}_{\pi}
}{m_{\rho}(m_{\rho}+E_{\rho})})-{B\over 3}{m_{a}\over m_{\rho}}
p^{2}_{\pi}}=-0.097.
\end{equation}
The quark model[24] predicts that \({d\over s}=-0.15.\)
Experimental value is $-0.11\pm 0.02$[25].

The vertex of $a_{1}\rightarrow\pi\gamma$ has been obtained by
using substitution(32) in eq.(61)
\begin{eqnarray}
{\cal L}_{a_{1}\rightarrow\pi\gamma}=-
\epsilon_{3jk}\{Aa^{\mu}_{j}A_{\mu}\pi_{k}
+Ba^{\mu}_{j}A^{\nu}\partial_{\mu\nu}\pi_{k}\},\nonumber \\
A={2\over f_{\pi}}(1-{1\over 2\pi^{2}g^{2}})^{-{1\over 2}}
\{{F^{2}\over g^{2}}+\frac{m^{2}_{a}}
{2\pi^{2}g^{2}}-[{2c\over g}+
\frac{3}{4\pi^{2}g^{2}}(1-{2c\over g})](m^{2}_{a}-q^{2})\},
\end{eqnarray}
where q is photon momentum and A is obtained from eq.(61).
Before presenting the numerical
result, let's prove the current conservation in the case of real
photon.
In order to have current conservation following equation should be
satisfied
\begin{equation}
A(q^{2}=0)={1\over 2}m^{2}_{a}B .
\end{equation}
The left hand of this equation can be written as
\begin{equation}
{2\over f_{\pi}}(1-{1\over 2\pi^{2}g^{2}})^{-{1\over 2}}\{
(1-{2c\over g})(1-{1\over 2\pi^{2}g^{2}})m^{2}_{a}-m^{2}_{\rho}
-{1\over 4\pi^{2}g^{2}}(1-{2c\over g})m^{2}_{a}\}.
\end{equation}
Using the mass formula(27), and the expression of c(25) it can be
found that the left hand of the equation(65) is just ${1\over 2}
m^{2}_{a}B$. Therefore, the current conservation in the process of
$a_{1}\rightarrow\pi\gamma$ is satisfied. The decay width of $a_{1}
\rightarrow\pi\gamma$ has been computed and is 252keV.
The experimental value is $640\pm 246$ keV[26].

Use ${\cal L}_{RE}$ the decay width of $a_{1}\rightarrow 3\pi$ can be
calculated. It is found that the branch ratio
is $5\times 10^{-4}$ which is consistent with data
$0.003\pm 0.003$[27].

{\large\bf Revisit of Weinberg's sum rules}\\
{}From chiral symmetry, current algebra, and VMD Weinberg has found the
first sum rule[4]
\begin{equation}
{g^{2}_{\rho}\over m^{2}_{\rho}}-{g^{2}_{a}\over m^{2}_{a}}={1\over
4}f^{2}_{\pi},
\end{equation}
where
\begin{eqnarray}
<0|\bar{\psi}{\tau_{i}\over 2}\gamma_{\mu}\psi|\rho^{j}_{\lambda}>
=g_{\rho}\epsilon^{\lambda}_{\mu},\nonumber \\
<0|\bar{\psi}{\tau_{i}\over 2}\gamma_{\mu}\gamma_{5}\psi|
\rho^{j}_{\lambda}>
=g_{a}\epsilon^{\lambda}_{\mu}.
\end{eqnarray}
Assuming an additional condition[4], Weinberg's second sum rule is
obtained
\begin{equation}
g_{a}=g_{\rho}.
\end{equation}
Eqs.(66,68) and KSFR sum rule together lead to \(m^{2}_{a}=2m^{2}
_{\rho}\) which is not in good agreement with present value of
$m_{a}$.
The theory presented in this paper can be considered as a realization
of chiral symmetry,
current algebra, and VMD. In this theory, the isovector vector and
axial-vector currents in eq.(1) are the same with the ones in eqs.(67).
Therefore, $g_{\rho}$ and $g_{a}$ can be evaluated
explicitly. Using eqs.(40,42) following expressions have been found
\begin{eqnarray}
<\bar{\psi}{\tau_{i}\over 2}\gamma_{\mu}\gamma_{5}\psi>=-{1\over 2}
gm^{2}_{\rho}(1-{1\over 2\pi^{2}g^{2}})^{-{1\over 2}}a^{i}_{\mu}
+... ,\nonumber \\
<\bar{\psi}{\tau_{i}\over 2}\gamma_{\mu}\psi>=-{1\over 2}
gm^{2}_{\rho}\rho^{i}_{\mu}+... .
\end{eqnarray}
We obtain
\begin{eqnarray}
g_{\rho}=-{1\over 2}gm^{2}_{\rho},\nonumber \\
g_{a}=-{1\over 2}gm^{2}_{\rho}(1-{1\over 2\pi^{2}g^{2}})^{-{1\over 2}}.
\end{eqnarray}
The relation(68) is not satisfied and \(m^{2}_{a}=2m^{2}_{\rho}\)
is not confirmed by this theory. This is the reason why
the mass relation(27) obtained in this paper is not the same with
the one of ref.[4]. However, Weinberg's first sum rule(66) only
depends on chiral symmetry, VMD, and current algebra. Therefore,
it should be achieved in this theory. Substituting
$g_{\rho}$ and $g_{a}$(70) into the left hand side of eq.(66),
we obtain
\begin{equation}
{g^{2}\over 4}m^{2}_{\rho}\{1-\frac{m^{2}_{\rho}}{m^{2}_{a}}
(1-{1\over 2\pi^{2}g^{2}})^{-1}\}.
\end{equation}
Substituting the mass formula(27) and eq.(26) into eq.(71), indeed
Weinberg's first sum rule is satisfied.
The factor $(1-{1\over 2\pi^{2}g^{2}})^{-{1\over 2}}$
plays important role in this theory.

In ref.[28]
two new formulas of the amplitude of $a_{1}\rightarrow
\rho\pi$ in the limit of $p_{\pi}\rightarrow 0$
have been found from the Ward identity found by Weinberg[4] and VMD
\begin{eqnarray}
g_{a}f_{\pi}A(m^{2}_{\rho})=2g_{\rho}(m^{2}_{a}-m^{2}_{\rho}),
\;\;\;g_{\rho}f_{\pi}A(m^{2}_{a})=2g_{a}(m^{2}_{a}-m^{2}_{\rho}).
\end{eqnarray}
It needs to check if
these two relations are satisfied in this theory. According to
ref.[28], in the limit
of \(p_{\pi}=0\) the amplitude A(61) of
$a_{1}\rightarrow\rho\pi$ can be written as
\begin{equation}
A(k^{2})={2\over f_{\pi}}\{\frac{F^{2}}{g^{2}}+\frac{k^{2}}{2\pi^{2}
g^{2}}\}(1-{1\over 2\pi^{2}g^{2}})^{-{1\over 2}}.
\end{equation}
Using eq.(73) and the mass formula(27), the two relations(72)
are indeed satisfied.

{\large\bf The decays of $\tau\rightarrow\rho\nu$ and $a_{1}\nu$} \\
The decay rates of $\tau\rightarrow\rho\nu$ and $a_{1}\nu$ can be
calculated in terms of the two matrix elements(67,70)
\begin{eqnarray}
\Gamma(\tau\rightarrow\rho\nu)=\frac{G^{2}}{8\pi}cos^{2}\theta
g^{2}_{\rho}\frac{m^{3}_{\tau}}{m^{2}_{\rho}}
(1-\frac{m^{2}_{\rho}}
{m^{2}_{\tau}})^{2}
(1+2\frac{m^{2}_{\rho}}{m^{2}_{\tau}})=4.84\times 10^{-13}GeV,
\nonumber \\
\Gamma(\tau\rightarrow a_{1}\nu)=\frac{G^{2}}{8\pi}cos^{2}\theta
g^{2}_{a}\frac{m^{3}_{\tau}}{m^{2}_{a}}
(1-\frac{m^{2}_{a}}
{m^{2}_{\tau}})^{2}
(1+2\frac{m^{2}_{a}}{m^{2}_{\tau}})=1.56\times 10^{-13}GeV.
\end{eqnarray}
The experimental values are $(0.495\pm 0.023)10^{-12}GeV$ and
$(2.42\pm 0.76)10^{-13}GeV$ for $\tau\rightarrow\rho\nu$ and
$\tau\rightarrow a_{1}\nu$ respectively.

These calculations can also be done in terms of the effective
lagrangian(13) of mesons, in which the couplings between the mesons and
W bosons are determined. In chiral limit, the $\pi-W$ coupling
has been found from the first term of the lagrangian(13)
\begin{equation}
-\frac{F^{2}}{4f_{\pi}}(1-\frac{2c}{g})g_{w}\partial_{\mu}\pi_{i}
W^{\mu}_{i}=-\frac{f_{\pi}}{4}g_{w}\partial_{\mu}\pi_{i}W^{\mu}_{i}.
\end{equation}
{}From $\pi_{l2}$ decay, $f_{\pi}$ has been determined to be 186MeV.

Like photon, from the lagrangian(1) it can be seen that W bosons
always appear in the combinations either
$\rho_{\mu}+
\frac{g_{w}}{4}g(\tau_{1}W^{1}_{\mu}+\tau_{2}W^{2}_{\mu})$ or
$a_{\mu}-
\frac{g_{w}}{4}g'_{a}(\tau_{1}W^{1}_{\mu}+\tau_{2}W^{2}_{\mu})$.
The W boson fields of eq.(1) are needed to be normalized.
\[W\rightarrow (1+2g^{2}{g^{2}_{w}\over 4})^{-{1\over 2}}W,\]
\[g_{w}\rightarrow (1+2g^{2}{g^{2}_{w}\over 4})^{{1\over 2}}
g_{w},\;\;\; g_{w}W_{\mu}\rightarrow g_{w}W_{\mu}.\]
Like VMD, the coupling of $\rho-W$ has been found
\begin{equation}
-{g_{w}\over 4}{1\over 2}g(\partial_{\mu}\rho^{i}_{\nu}-\partial_{\nu}
\rho^{i}_{\mu})(\partial^{\mu}W^{i\nu}-\partial^{\nu}W^{i\mu}).
\end{equation}
When $\rho$ is on mass shell the coupling becomes $g_{\rho}$.
Of course, like VDM, there is direct coupling between W boson and
other mesons
\begin{equation}
{g_{w}\over 4}{1\over 2}gW^{i}_{\mu}j^{i\mu}.
\end{equation}
The axial-vector part of the interactions
of W boson with mesons has been found
\begin{eqnarray}
-{g_{w}\over 4}g(1-\frac{1}{2\pi^{2}g^{2}})^{-{1\over 2}}(\frac{F^{2}
}{g^{2}}+{1\over 2\pi^{2}g^{2}}m^{2}_{a})a^{i}_{\mu}W^{i\mu}
-{g_{w}\over 4}f_{\pi}W_{i\mu}\partial_{\mu}\pi_{i}
\nonumber \\
+{g_{w}\over 4}{1\over 2}g(1-\frac{1}{2\pi^{2}g^{2}})^{-{1\over 2}}
(\partial_{\mu}a^{i}_{\nu}-\partial_{\nu}
a^{i}_{\mu})(\partial^{\mu}W^{i\nu}-\partial^{\nu}W^{i\mu})
-{g_{w}\over 4}{1\over 2}gW^{i}_{\mu}j^{i\mu}_{a},
\end{eqnarray}
where $j^{i\mu}_{a}$ is defined as isovector axial-vector current
and $a_{1}$ fields couple to this current.
Using the mass formula(27), it can be seen from eq.(78) that
the coupling $a_{1}-W$ is just $g_{a}$(70).

{\large\bf $\pi\pi$ scattering and determination of parameters of
CPT}\\
The pion is nearly Goldstone boson associated with the dynamically
broken $SU(2)_{L}\times SU(2)_{R}$ chiral symmetry which is a symmetry
of $QCD$ in the limit $m_{u,d}\rightarrow 0$. $\pi\pi$ scattering
has been studied by Weinberg, using nonlinear chiral lagrangian[29].
The modern study of $\pi\pi$ scattering utilizes chiral
perturbation theory[2,30]. In present theory the $\pi\pi$ scattering
is related to even number of $\gamma_{5}$, hence the lagrangian
of this process can be found from eq.(13). As discussed in ref.[31]
the pion mass term can be introduced to the theory by adding
the quark mass term $\bar{\psi}M\psi$(M is the quark mass matrix)
to the lagrangian(1). According to ref.[31] the pion mass term
obtained from the quark mass term is
\begin{equation}
\frac{1}{8}f^{2}_{\pi}m^{2}_{\pi}Tr(U-1).
\end{equation}
The $\pi_{j_{1}}(k_{1})+\pi_{j_{2}}(k_{2})\rightarrow\pi_{i_{1}}
(p_{1})\pi_{i_{2}}(p_{2})$ scattering amplitudes are written as
\begin{equation}
T_{j_{1}j_{2},i_{1}i_{2}}=A(s,t,u)\delta_{i_{1}i_{2}}\delta_{j_{1}
j_{2}}+A(t,s,u)\delta_{i_{1}j_{1}}\delta_{i_{2}j_{2}}+A(u,t,s)
\delta_{i_{1}j_{2}}\delta_{i_{2}j_{1}},
\end{equation}
where \(s=(k_{1}+k_{2})^{2}\), \(t=(k_{1}-p_{1})^{2}\),
\(u=(k_{1}-p_{2})^{2}\). In the frame of center of mass \(s=4m^{2}
_{\pi}+4k^{2}\), \(t=-2k^{2}(1-cos\theta)\),
\(u=-2k^{2}(1+cos\theta)\),
where k is the pion momentum and $\theta$ is the scattering angle.
The partial wave amplitudes are defined as
\begin{eqnarray}
T^{I}_{l}(s)=\frac{1}{64\pi}\int^{1}_{-1}dcos\theta P_{l}(cos\theta)
T^{I}(s,t,u),\nonumber \\
T^{0}=3A(s,t,u)+A(t,s,u)+A(u,t,s),\nonumber \\
T^{1}=A(t,s,u)-A(u,t,s),\nonumber \\
T^{2}=A(t,s,u)+A(u,t,s).
\end{eqnarray}
At low energies, the partial wave amplitudes can be expanded in
terms of scattering length $a^{I}_{l}$ and slope $b^{I}_{l}$
\begin{equation}
ReT^{I}_{l}(s)=(\frac{k^{2}}{m^{2}_{\pi}})^{l}(a^{I}_{l}+{k^{2}
\over m^{2}_{\pi}}b^{I}_{l}).
\end{equation}
The lagrangian of $\pi\pi$ scattering derived from eq.(13) contains
two parts: direct coupling and $\rho$ meson exchange.
To the leading order of chiral perturbation,
the amplitudes obtained from direct coupling(with an index D) are
\begin{eqnarray}
\lefteqn{A(s,t,u)_{D}=\frac{16}{f^{4}_{\pi}}\{{1\over 3}f^{2}_{\pi}
(1-\frac{6c}{g})(2m^{2}_{\pi}+3k^{2})
+{c^{2}\over g^{2}}({3\over \pi^{2}}(1-\frac{2c}{g})^{2}-4c^{2})
(6k^{4}-2k^{4}cos^{2}\theta)}\nonumber \\
& &-\frac{4}{(4\pi)^{2}}(1-{2c\over g})^{4}(10k^{4}
-2k^{4}cos^{2}\theta)+
\frac{8}{(4\pi)^{2}}(1-{2c\over g})^{2}[-{16c\over g}
k^{4}+4k^{4}+{4c^{2}\over g^{2}}k^{4}(1+cos^{2}\theta)]\},
\nonumber \\
&&A(t,s,u)_{D}=\frac{16}{f^{4}_{\pi}}\{{1\over 6}f^{2}_{\pi}(1-
{6c\over g})(-2m^{2}_{\pi}-3k^{2}+3k^{2}cos\theta)\nonumber \\
& &+{c^{2}\over g^{2}}(\frac{3}{\pi^{2}}(1-{2c\over g})^{2}-4c^{2})
[-3k^{4}+k^{4}cos^{2}\theta-6k^{4}cos\theta] \nonumber \\
& &-{4\over (4\pi)^{2}}
(1-{2c\over g})^{4}[-2k^{4}+2k^{4}cos^{2}\theta
-8k^{4}cos\theta]\nonumber \\
& &+\frac{8}{(4\pi)^{2}}(1-{2c\over g})^{2}[{2c\over g}
(-2k^{4}-2k^{4}cos^{2}\theta+4k^{4}cos\theta)
+(-k^{2}+k^{2}cos\theta)^{2}\nonumber \\
& &+{2c^{2}\over g^{2}}(5k^{4}+2
k^{4}cos\theta+k^{4}cos^{2}\theta)]\}.
\end{eqnarray}
The eq.(26) has been used in deriving eqs.(83).
The amplitude A(u,t,s) can be obtained by using the substitution of
$cos\theta\rightarrow -cos\theta$ in A(t,s,u).
The amplitudes from $\rho$ exchange have been obtained by using
eqs.(48,49)
\begin{eqnarray}
A(s,t,u)_{\rho}={8\over g^{2}}\frac{2m^{2}_{\pi}+3k^{2}+k^{2}
cos\theta}{m^{2}_{\rho}+2k^{2}-2k^{2}cos\theta}+{8\over g^{2}}
\frac{2m^{2}_{\pi}+3k^{2}-k^{2}cos\theta}{m^{2}_{\rho}+2k^{2}
+2k^{2}cos\theta},\nonumber \\
A(t,s,u)_{\rho}={16\over g^{2}}\frac{k^{2}cos\theta}{m^{2}
_{\rho}-s+im_{\rho}\Gamma(k)}-{8\over g^{2}}\frac{2m^{2}_{\pi}
+3k^{2}-k^{2}cos\theta}{m^{2}_{\rho}+2k^{2}+2k^{2}cos\theta},
\nonumber \\
A(u,t,s)_{\rho}={-16\over g^{2}}\frac{k^{2}cos\theta}{m^{2}
_{\rho}-s+im_{\rho}\Gamma(k)}-{8\over g^{2}}\frac{2m^{2}_{\pi}
+3k^{2}+k^{2}cos\theta}{m^{2}_{\rho}+2k^{2}-2k^{2}cos\theta},
\end{eqnarray}
where $\Gamma(k)$ is the decay width of $\rho$ meson
\begin{equation}
\Gamma(k)={2\over 3\pi g^{2}}\frac{k^{3}}{m^{2}_{\rho}}.
\end{equation}
Due to kinematic reason the decay width of $\rho$ meson
appears only when the virtual momentum squared of $\rho$
is equal to s.

The scattering lengths and slopes have been found from direct
coupling(eqs.(83))
\begin{eqnarray}
a^{0}_{0}=\frac{5m^{2}_{\pi}}{24\pi f^{2}_{\pi}}+\frac{2m^{2}_
{\pi}}{3\pi f^{2}_{\pi}}(1-{6c\over g}),\nonumber \\
b^{0}_{0}=\frac{m^{2}_{\pi}}{\pi f^{2}_{\pi}}(1-{6c\over g}),
\nonumber \\
a^{2}_{0}=\frac{m^{2}_{\pi}}{12\pi f^{2}_{\pi}}-\frac{m^{2}
_{\pi}}{3\pi f^{2}_{\pi}}(1-{6c\over g}),\nonumber \\
b^{2}_{0}=-\frac{m^{2}_{\pi}}{2\pi f^{2}_{\pi}}(1-{6c\over g}),
\nonumber \\
a^{1}_{1}=\frac{m^{2}_{\pi}}{6\pi f^{2}_{\pi}}
(1-{6c\over g}),
\end{eqnarray}
and from $\rho$ exchange(84) we obtain
\begin{eqnarray}
a^{0}_{0}=\frac{2m^{2}_{\pi}}{\pi g^{2}m^{2}_{\rho}},\;\;
b^{0}_{0}=\frac{3m^{2}_{\pi}}{\pi g^{2}m^{2}_{\rho}},\;\;
a^{2}_{0}=-\frac{m^{2}_{\pi}}{\pi g^{2}m^{2}_{\rho}},\nonumber \\
b^{2}_{0}=-\frac{3m^{2}_{\pi}}{2\pi g^{2}m^{2}_{\rho}},\;\;
a^{1}_{1}=\frac{m^{2}_{\pi}}{2\pi g^{2}m^{2}_{\rho}}.
\end{eqnarray}
Numerical calculation shows that in these quantities the
contribution of $\rho$ exchange is dominant, for instance,
the contribution of $\rho$ exchange to $a^{0}_{0}$ is ten more
times of the one from direct coupling. Adding eqs.(86,87) together
and using eq.(25) we obtain
\begin{eqnarray}
a^{0}_{0}=\frac{7m^{2}_{\pi}}{8\pi f^{2}_{\pi}},\;\;
b^{0}_{0}=\frac{m^{2}_{\pi}}{\pi f^{2}_{\pi}},\;\;
a^{2}_{0}=-\frac{m^{2}_{\pi}}{4\pi f^{2}_{\pi}},\;\;
b^{2}_{0}=-\frac{m^{2}_{\pi}}{2\pi f^{2}_{\pi}},\;\;
a^{1}_{1}=\frac{m^{2}_{\pi}}{6\pi f^{2}_{\pi}}.
\end{eqnarray}
These are just the scattering lengths and slopes obtained by
Weinberg[29]. In eqs.(86), there are terms with the factor of
${c\over g}$ obtained from the shift $a_{\mu}\rightarrow a_{\mu}
-c\partial_{\mu}\pi$. Due to eq.(25) these terms are cancelled by
the corresponding terms obtained from $\rho$ exchange. These
cancellations result Weinberg's results in this theory.
As a matter of fact, the cancellation is the result of
chiral symmetry. In the lagrangian(1), there is a term ${1\over 2}
m^{2}_{0}(a_{\mu}a^{\mu}+v_{\mu}v_{\mu})$ introduced by chiral
symmetry and due to this term, c(see eq.(25)) has $m^{2}_{\rho}$ in
the denominator of the expression(25) which leads to the cancellation.
All these quantities are only related to the zeroth and the
second orders of derivatives. The scattering lengths and slope
$a^{0}_{2}$, $a^{2}_{0}$, and $b^{1}_{1}$ have been found from the
terms with the derivatives at the fourth order
\begin{eqnarray}
b^{1}_{1}=\frac{m^{4}_{\pi}}{6\pi^{3}f^{4}_{\pi}}(-0.036)+\frac{
2m^{4}_{\pi}}{\pi g^{2}m^{4}_{\rho}},\nonumber \\
a^{0}_{2}=\frac{m^{4}_{\pi}}{10\pi^{3}f^{4}_{\pi}}(0.0337)+
\frac{4m^{4}_{\pi}}{15\pi g^{2}m^{4}_{\rho}},\nonumber \\
a^{2}_{2}=\frac{m^{4}_{\pi}}{10\pi^{3}f^{4}_{\pi}}(0.0207)-
\frac{2m^{4}_{\pi}}{15\pi g^{2}m^{4}_{\rho}}.
\end{eqnarray}
In these quantities(89), the contributions of $\rho$ exchange
are ten more times of the ones from direct coupling. $\rho$
meson exchange is dominant. The numerical results are listed
in Table I.
\begin{table}[h]
\begin{center}
\caption{Table I The pion scattering lengths and slopes}
\begin{tabular}{|c|c|c|} \hline
    &  Experimental  &  Theoretical  \\ \hline
$a^{0}_{0}$ & $0.26\pm 0.05$ & 0.16 \\ \hline
$b^{0}_{0}$ & $0.25\pm 0.03$ & 0.18 \\ \hline
$a^{2}_{0}$ & $-0.028\pm 0.012$ & -0.045 \\ \hline
$b^{2}_{0}$ & $-0.082\pm 0.008$ & -0.089 \\ \hline
$a^{1}_{1}$ & $0.038\pm 0.002$  & 0.030  \\ \hline
$b^{1}_{1}$ &                 & $5.56\times 10^{-3}$ \\ \hline
$a^{0}_{2}$ & $(17\pm 3)\times 10^{-4}$ & $7.84\times 10^{-4}$ \\
\hline
$a^{2}_{2}$ & $(1.3\pm 3)\times 10^{-4}$ & $-3.53\times 10^{-4}$ \\
\hline
\end{tabular}
\end{center}
\end{table}

It is well known that the chiral perturbation theory[2,30] describes
$\pi\pi$ scattering pretty well. In principle the parameters of
chiral perturbation theory can be calculated by present theory.
Besides $f_{\pi}$ and $m_{\pi}$ there are other two parameters
appearing in $\pi\pi$ scattering[30]
\begin{equation}
{\cal L}_{4}={\alpha_{1}\over 4}Tr\{\partial_{\mu}U\partial^{\mu}
U^{\da}\}^{2}+{\alpha_{2}\over 4}Tr(\partial_{\mu}U\partial_{\nu}
U^{\da})Tr(\partial^{\mu}U\partial^{\nu}U^{\da}).
\end{equation}
The amplitudes $T^{0}_{2}$ and $T^{2}_{2}$ have been determined
by ${\cal L}_{4}$[30]
\begin{equation}
T^{0}_{2}=\frac{2\alpha_{2}+\alpha_{1}}{15\pi f^{4}_{\pi}}
(s-4m^{2}_{\pi})^{2},\;\;\;
T^{2}_{2}=\frac{2\alpha_{1}+\alpha_{2}}{30\pi f^{4}_{\pi}}
(s-4m^{2}_{\pi})^{2}.
\end{equation}
{}From the amplitudes(83,84)(to $O(k^{4})$), we obtain
\begin{equation}
T^{0}_{2}=\frac{1}{15\pi f^{4}_{\pi}}(s-4m^{2}_{\pi})^{2}
(0.00698),\;\;\;T^{2}_{2}=-\frac{1}{30\pi f^{4}_{\pi}}
(s-4m^{2}_{\pi})^{2}(0.00656).
\end{equation}
The two parameters in eq.(91) have been determined
\begin{equation}
\alpha_{1}=-0.0068,\;\;\;\alpha_{2}=0.0070.
\end{equation}
They are compatible with the values of CPT[30]
\begin{equation}
\alpha_{1}=-0.0092,\;\;\;\alpha_{2}=0.0080.
\end{equation}
The same values of $\alpha_{1}$ and $\alpha_{2}$ can also be found
by comparing $T^{0}_{0}$, $T^{2}_{0}$, and $T^{1}_{1}$ of ref.[30]
with the corresponding combinations of eqs.(83,84). The numerical
calculation shows that the contribution of $\rho$-exchange to
$\alpha_{1}$ and $\alpha_{2}$ is higher than the one from
direct coupling by two order of magnitude. The contribution
of $\rho$-exchange to $T^{0}_{2}$ and $T^{2}_{2}$ can be obtained
from eqs.(84)
\[T^{0}_{2}=\frac{4}{15\pi g^{2}m^{4}_{\rho}}k^{4},\;\;\;
T^{2}_{2}=-\frac{2}{15\pi g^{2}m^{4}_{\rho}}k^{4}.\]
Then we have
\[-\alpha_{1}=\alpha_{2}={1\over 4}{1\over g^{2}}{f^{4}_{\pi}\over
m^{4}_{\rho}}=0.007.\]
Using the decay width of $\rho$ meson(85), we obtain
\[\alpha_{1}=-\alpha_{2}=-3\pi{f^{4}_{\pi}\over m^{5}_{\rho}}
\frac{\Gamma_{\rho}}{(1-\frac{4m^{2}_{\pi}}{m^{2}_{\rho}})^{2\over 3}
}.\]
This is just the expression presented in ref.[8]. In present theory
the reason of $\rho$ dominance is the consequence of the
cancellation between the original pion and the pion obtained from
the shift of $a_{\mu}\rightarrow a_{\mu}-c\partial_{\mu}\pi$.
On the other hand,
the energy dependence of the amplitudes(83,84) have been predicted
by present theory.
The amplitudes $A(t,s,u)_{\rho}$ and $A(u,t,s)_{\rho}$
(84) predict the resonance structure of $\rho(770)$ in the
channel of \(I=1\) and \(l=1\). The experimental data[32]
clearly shows the $\rho(770)$ resonance in the amplitude
$T^{1}_{1}$. The comparison between theoretical predictions
and experimental data are shown in Fig.1.
{}From Fig.1, it can be seen that $|T^{1}_{1}|$ is in good agreement
with data and $T^{0}_{2}$ ,$ReT^{2}_{2}$, and $ReT^{2}_{0}$
agree with data well. However, this theory does not provide an
imaginary part for $T^{0}_{0}$ and the experimental data shows
that there is an imaginary part in $T^{0}_{0}$. On the other
hand, the data shows(Table I) that the theoretical predictions of
$a^{0}_{0}$ and $b^{0}_{0}$ are lower than the experimental
data. Therefore, in the channel of \(I=0\) and \(l=0\) something
is missing in this theory. In ref.[33] the $0^{++}$ $f_{0}(1300)$
meson has been introduced to the effective meson theory to
improve the theoretical value of $a^{0}_{0}$. The $f_{0}(1300)$
meson can be introduced to the lagrangian(1), however, this is
beyond the scope of this paper.

{\large\bf $\omega\rightarrow \rho\pi$ and other related processes}\\
In 60's, Gell-Mann, Sharp and Wagner[34] used the coupling of $\omega
\rightarrow\rho\pi$ to compute the decay rates of $\omega\rightarrow
\pi\gamma$ and $\pi^{0}\rightarrow\gamma\gamma$ in terms of VMD.
The Syracuse group[12] has
used the generalized Wess-Zumino action
to study these processes.

In the process $\omega\rightarrow\rho\pi$ only pion is associated with
$\gamma_{5}$ in this theory.
Therefore, this process should be described by the
effective lagrangian ${\cal L}_{IM}$(6).
In ref.[20] we use the lagrangian without vector and
axial-vector
mesons to calculate the effective baryon current
${1\over 3}<\bar{\psi}\gamma_{\mu}\psi>$ and it has been found that
the current found in ref.[20] is just the topological current induced
from the Wess-Zumino term. In this theory the interactions of $\omega$
meson with others are through the term ${1\over g}\omega^{\mu}
<\bar{\psi}\gamma_{\mu}\psi>$ in the lagrangian(1).
It is the same with ref.[20] that
the current $<\bar{\psi}\gamma_{\mu}\psi>$ can be bosonized by
following equation
\begin{equation}
<\bar{\psi}\gamma_{\mu}\psi>=\frac{-i}{(2\pi)^{D}}\int d^{D}p
Tr\gamma_{\mu}s_{F}(x,p).
\end{equation}
The leading terms come from \(n=3\).
After a lengthy and very careful derivation
it has been found that except the term $g\partial^{2}\omega_{\mu}$
in eq.(45)
all other terms with even number of $\gamma_{5}$
in eq.(95)
are cancelled each other. This is consistent with the fact that
in two flavor case $\omega$ field is a
flavor singlet except the kinetic term, $\omega$ field
does not appear in ${\cal L}_{RE}$(13)
which describes the processes with normal parity.
Using eq.(42) and taking \(n=3\), we have
\begin{equation}
<\bar{\psi}\gamma_{\mu}\psi>=\frac{i}{(2\pi)^{D}}\int\frac{d^{D}p
}{(p^{2}-m^{2})^{4}}Tr\gamma_{\mu}(\sspp-m\hat{u})\ssD(\sspp-m\hat
{u})\ssD(\sspp-m\hat{u})\ssD(\sspp-m\hat{u}),
\end{equation}
where \(\ssD=i\ssp+{1\over g}\ssv+{1\over g}(1-\frac{1}{2\pi^{2}
g^{2}})^{-{1\over 2}}\ssa\gamma_{5}\).
In the calculation of eq.(96) we use dimension regularization.
However, there is one term in eq.(96)
\begin{equation}
\frac{i}{(2\pi)^{D}}\int\frac{d^{D}p}{(2\pi)^{D}}Tr\gamma_{\mu}
\sspp\ssD\sspp\ssD\sspp\ssD\sspp
\end{equation}
which has divergence and contains
$\gamma_{5}$. Therefore this part of the integral needs special
treatment in dimension regularization. We use t`Hooft and Veltman's
prescription[35] used to treat the triangle anomaly to calculate this
integral. We define
\[p=\underline{p}+q\]
where $\underline{p}$ has only four components in the 4-dimensional
space and $q$ is defined in $D-4$ dimensional space. According to
ref.[35] $\underline{p}$ and q observe following equations
\[\ssun\gamma_{5}=-\gamma_{5}\ssun,\;\;\;
\ssq\gamma_{5}=\gamma_{5}\ssq.\]
Following these treatments the integral has been computed. The final
expression of \\
$<\bar{\psi}\gamma_{\mu}\psi>$ is
\begin{eqnarray}
\lefteqn{{1\over g}\omega^{\mu}<\bar{\psi}\gamma_{\mu}\psi>=
g\partial^{2}\omega} \nonumber \\
& &+\frac{N_{c}}{(4\pi)^{2}g}\varepsilon^{\mu\nu\alpha\beta}
\omega_{\mu}Tr
\{{4\over 3g^{2}}[-v_{\nu\alpha}a_{\beta}+\partial_{\nu}(a_{\alpha}
v_{\beta})-ia_{\nu}a_{\alpha}a_{\beta}-\frac{i}{2}a_{\alpha}
[\rho_{\beta},\rho_{\nu}]]\nonumber \\
& &-{2\over 3}(V_{\nu\alpha}a_{\beta}
-a_{\nu\alpha}V_{\beta})
-{1\over 3}(V_{\nu\alpha}a_{\beta}+a_{\nu\alpha}V_{\beta})\}
\nonumber \\
 & &+\frac{N_{c}}{(4\pi)^{2}g}{i\over 3}
\varepsilon^{\mu\nu\alpha\beta}\omega_{\mu}
Tr\{2(D_{\nu}U)U^{\da}(D_{\alpha}U)U^{\da}(D_{\beta}U)U^{\da}
-{3i\over g^{2}}V_{\nu\alpha}a_{\beta}-{3i\over g^{2}}
a_{\nu\alpha}\rho_{\beta}\nonumber \\
& &+{3i\over g}V_{\nu\alpha}[U(D_{\beta}U^{\da})-U^{\da}(D_{\beta}U)]
-{3i\over g}a_{\nu\alpha}[U(D_{\beta}U^{\da})+U^{\da}(D_{\beta}U)]\}
\end{eqnarray}
where \(v_{\nu\alpha}=\partial_{\nu}\rho_{\alpha}-\partial_{\alpha}
\rho_{\nu}-{i\over g}[\rho_{\nu},\rho_{\alpha}]\),
\(V_{\nu\alpha}=\partial_{\nu}\rho_{\alpha}-\partial_{\alpha}
\rho_{\nu}-{i\over g}[\rho_{\nu},\rho_{\alpha}] -{i\over g}
[a_{\nu},a_{\alpha}]\), and \(a_{\nu}\rightarrow (1-{1\over
2\pi^{2}g^{2}}
)^{-{1\over 2}}a_{\nu}-c\partial_{\nu}\pi\).
This expression can be simplified as
\begin{eqnarray}
\lefteqn{{1\over g}\omega^{\mu}<\bar{\psi}\gamma_{\mu}\psi>=
\frac{N_{c}}{(4\pi)^{2}g}{2\over 3}\varepsilon^{\mu\nu\alpha\beta}
\omega_{\mu}Tr\partial_{\nu}UU^{\da}\partial_{\alpha}UU^{\da}
\partial_{\beta}UU^{\da}}\nonumber \\
 & &+\frac{N_{c}}{(4\pi)^{2}}{2\over g}\varepsilon^{\mu\nu\alpha\beta}
\partial_{\mu}\omega_{\nu}Tr\{{i\over g}[\partial_{\beta}UU^{\da}
(\rho_{\alpha}+a_{\alpha})-\partial_{\beta}U^{\da}U(\rho_{\alpha}
-a_{\alpha})]\nonumber \\
& &-{2\over g^{2}}(\rho_{\alpha}+a_{\alpha})U(\rho_{\beta}-a_{\beta})
U^{\da}-{2\over g^{2}}\rho_{\alpha}a_{\beta}\}.
\end{eqnarray}
This formula is exact the same with the one obtained by Syracuse
group[12]. In ref.[12] the Bardeen form of the anomaly[36] has been
accepted and an arbitrary constant in their formula has
been chosen to be 1. The authors of ref.[12] claim that their
Wess-Zumino lagrangian with spin-1 fields agrees with Witten's[11]
expression except an inadvertently omitted term.
In eq.(99) all the couplings are fixed by g and c. This is the
universality of coupling in this theory.

The interaction lagrangian of $\omega\rho\pi$ has been found from
eq.(99)
\begin{equation}
{\cal L}_{\omega\rho\pi}=-\frac{N_{c}}{\pi^{2}g^{2}f_{\pi}}
\varepsilon^{\mu\nu\alpha\beta}\partial_{\mu}\omega_{\nu}
\rho^{i}_{\alpha}\partial_{\beta}\pi^{i},
\end{equation}
which can be used to study related processes.
Using the vertex of $\omega\rightarrow \pi\gamma$ obtained by
the substitution $\rho^{0}_{\mu}\rightarrow {e\over f_{\rho}}
A_{\mu}$ in eq.(100) the decay width has been calculated
to be
\[\Gamma(\omega\rightarrow\gamma\pi)=724keV.\]
The experimental value is $717(1\pm 0.07)$keV. In the same way,
we obtain the vertex of
$\rho\rightarrow\pi\gamma$ by the substitution
$\omega_{\mu}\rightarrow{e\over f_{\omega}}A_{\mu}$ in eq.(100). The
decay width calculated is
\[\Gamma(\rho^{0}\rightarrow\pi^{0}\gamma)=76.2keV.\]
The experimental data is $68.2(1\pm 0.12)keV$.

$\pi^{0}\rightarrow\gamma\gamma$ is an evidence of the
Adler-Bell-Jackiw anomaly[37]. This process
is a crucial test of present theory. According to VMD, the
$\pi^{0}\gamma\gamma$ vertex should be obtained by
using the substitutions(32) in eq.(100).
\begin{equation}
{\cal L}_{\pi^{0}\rightarrow\gamma\gamma}=-{\alpha\over \pi
f_{\pi}}\varepsilon^{\mu\nu\alpha\beta}\pi^{0}\partial_{\mu}
A_{\nu}\partial_{\alpha}A_{\beta},
\end{equation}
which is the expression given by Adler-Bell-Jackiw anomaly[37].
The decay width obtained from eq.(101) is
\begin{equation}
\Gamma(\pi^{0}\rightarrow\gamma\gamma)=\frac{\alpha^{2}}{16\pi^{3}}
\frac{m^{3}_{\pi}}{f^{2}_{\pi}},
\end{equation}
which is the result of triangle anomaly.
The numerical result is 7.64eV and the data
is $7.74(1\pm 0.072)eV$.
If one of the two photons $\pi^{0}\rightarrow\gamma\gamma$ is
virtual, according to VMD, there are vector meson poles in the
decay amplitude.
The amplitude has been written as
\begin{equation}
{\cal M}=\frac{2\alpha}{\pi f_{\pi}}\{1+{1\over 2}\frac{q^{2}}{m^{2}_
{\rho}-q^{2}}+{1\over 2}\frac{q^{2}}{m^{2}_{\omega}-q^{2}}\},
\end{equation}
where q is the momentum of the virtual photon. Due to $q^{2}$ is
much less than the mass of the vector mason, the form factor
takes following form approximately
\begin{eqnarray}
F(q^{2})=1+{1\over 2}\frac{q^{2}}{m^{2}_{\rho}-q^{2}}
+{1\over 2}\frac{q^{2}}{m^{2}_{\omega}-q^{2}}=1+a\frac{q^{2}}{m^{2}_
{\pi^{0}}},\nonumber \\
a={m^{2}_{\pi^{0}}\over 2}({1\over m^{2}_{\omega}}+{1\over m^{2}
_{\rho}})=0.03.
\end{eqnarray}
The data is \( a=0.032\pm 0.004\).

In $\omega\rightarrow\pi\pi\pi$
besides the process $\omega\rho\pi$ and $\rho\rightarrow\pi\pi$
there is direct coupling $\omega\pi\pi\pi$
which has been derived from eq.(99)
\begin{equation}
{\cal L}_{\omega\pi\pi\pi}={2\over g\pi^{2}f^{3}_{\pi}}
(1+{6c^{2}\over g^{2}}-{6c\over g})\varepsilon^{\mu\nu\alpha\beta}
\varepsilon_{ijk}\omega_{\mu}\partial_{\nu}\pi_{i}\partial_{\alpha}
\pi_{j}\partial_{\beta}\pi_{k}.
\end{equation}
In both the ${\cal L}_{\omega\pi\pi\pi}$ and
${\cal L}_{\omega\rightarrow\rho\pi}$ there are a factor of
${1\over \pi^{2}}$. Therefore, qualitatively speaking, this
theory predicts a narrower width for $\omega$ decay.
Using the formula of c(25) and the value of g it is found that
\[1+{6c^{2}\over g^{2}}-{6c\over g}=-0.083.\]
The last two terms come from the shift
$a_{\mu}\rightarrow a_{\mu}-c\partial_{\mu}\pi$ and there is very
strong cancellation. Using eqs.(49,100,105) the decay width has
been obtained
\begin{eqnarray}
\Gamma(\omega\rightarrow\pi\pi\pi)=\frac{1}{24m_{\omega}(2\pi)^{3}}
\int dq^{2}_{1}dq^{2}_{2}\{|\vec{p}_{1}|^{2}|\vec{p}_{2}|^{2}-
(\vec{p}_{1}\cdot\vec{p}_{2})^{2}\}\nonumber \\
\{3f_{\omega 3\pi}
+f_{\omega\rho\pi}f_{\rho\pi\pi}({1\over q^{2}_{1}-m^{2}_{\rho}}+
{1\over q^{2}_{2}-m^{2}_{\rho}}+{1\over q^{2}_{3}-m^{2}_{\rho}})\},
\end{eqnarray}
where
\begin{equation}
f_{\omega 3\pi}={2\over g\pi^{2}f^{3}_{\pi}}(1+{6c^{2}
\over g^{2}}-{6c\over g}),\;\; f_{\omega\rho\pi}={N_{c}\over
\pi^{2}g^{2}f_{\pi}},
\end{equation}
and \(q^{2}_{i}=(p-p_{i})^{2}\), p is $\omega$ momentum and $p_{i}$
is pion momentum. We obtain
\[\Gamma(\omega\rightarrow 3\pi)=5 MeV.\]
If only $\omega\rightarrow\rho\pi$ and $\rho\rightarrow \pi\pi$ are
taken into account the width of $\omega\rightarrow 3\pi$ is 5.4MeV.
The experimental value is $7.43(1\pm 0.02)MeV$. From this study
we can see that the process of $\omega\rightarrow\rho\pi$ is dominant
the decay of $\omega\rightarrow 3\pi$ as proposed by the authors[34].
The direct coupling of $\omega\rightarrow 3\pi$ is responsible for
about $20\%$ of the decay rate.
The agreement between theoretical and experimental decay rates of
$\pi^{0}\rightarrow\gamma\gamma$, $\omega\rightarrow
\pi\gamma$, and $\rho\rightarrow\pi\gamma$ shows that
$\omega\rightarrow
\rho\pi$ obtained in this theory is more reliable.
However, at tree level due to the
cancellation  $f_{\omega3\pi}$ is too small and has a wrong sign.
Therefore, corrections from loop diagrams and terms with higher order
derivatives to $f_{\omega 3\pi}$ are needed.

{\large\bf The decays of $f_{1}(1285)$ meson}\\
In this theory $f_{1}(1285)$ meson is the chiral partner of
$\omega$ meson(see eq.(1)). The decay of $f_{1}\rightarrow 4\pi$
consists of two processes: direct coupling
$f_{1}\rightarrow \rho\pi\pi$
and $f_{1}\rightarrow a_{1}\pi\rightarrow
\rho\pi\pi$. $f_{1}$ meson is associated with $\gamma_{5}$
(see eq.(1)), hence, odd number of $\gamma_{5}$ are involved
in these two processes. Therefore, the decay of $f_{1}\rightarrow
\rho\pi\pi$
can not be found in eq.(13) and should be found in the Wess-Zumino
lagrangian with vector and axial-vector mesons. In this paper
the method used to find the effective
lagrangian of these processes is the same with the one used
to study the decays of $\omega$ meson. From the lagrangian(1)
and normalization of $f_{1}$ meson the vertices of $f_{1}$
decays can be found from following formula
\begin{equation}
{\cal L}_{i}={1\over g}(1-{1\over 2\pi^{2}g^{2}})^{-{1\over 2}}
f_{\mu}<\bar{\psi}\gamma_{\mu}\gamma_{5}\psi>.
\end{equation}
The bosonization of $<\bar{\psi}\gamma_{\mu}\gamma_{5}\psi>$
can be carried out by using following equation
\begin{equation}
<\bar{\psi}\gamma_{\mu}\gamma_{5}\psi>=\frac{-i}{(2\pi)^{D}}
\int d^{D}p Tr\gamma_{\mu}\gamma_{5}s_{F}(x,p).
\end{equation}
Substituting eq.(42) into eq.(109) the flavor singlet
axial-vector current
of mesons can be achieved. Due to the fact that only
odd number of $\gamma_{5}$ is involved in $f_{1}\rightarrow
\rho\pi\pi$, we
are only interested in the terms with the antisymmetric tensor.
The leading terms with antisymmetric tensor appear at \(n=3\).
It is similar to the case of $\omega$ meson there are divergent terms
which can be treated by the prescription provided in ref.[35].
Finally, the terms with $\varepsilon^{\mu\nu\alpha\beta}$ in
effective lagrangian(108) take following form
\begin{eqnarray}
\lefteqn{{\cal L}_{i}=-\frac{N_{c}}{3g^{3}(4\pi)^{2}}
(1-{1\over 2\pi^{2}g^{3}})
^{-{1\over 2}}\varepsilon^
{\mu\nu\alpha\beta}f_{\mu}Tr(3V_{\nu\alpha}V_{\beta}-a_{\nu\alpha}
a_{\beta})} \nonumber \\
 & &+\frac{iN_{c}}{3g^{2}(4\pi)^{2}}(1-{1\over 2\pi^{2}g^{2}})
 ^{-{1\over 2}}
\varepsilon^{\mu\nu\alpha\beta}f_{\mu}Tr\{-\partial_{\alpha}(\partial
_{\nu}UU^{\da}d^{+}_{\beta}+\partial_{\nu}U^{\da}Ud^{-}_{\beta})
\nonumber \\
& &-2(\partial_{\nu}UU^{\da}\partial_{\alpha}UU^{\da}d^{+}_{\beta}
+\partial_{\nu}U^{\da}U\partial_{\alpha}U^{\da}Ud^{-}_{\beta})
\nonumber \\
& &-{i\over g}(d^{+}_{\nu}\partial_{\alpha}d^{+}_{\beta}+d^{-}_{\nu}
\partial_{\alpha}d^{-}_{\beta}+2U^{\da}d^{+}_{\nu}U\partial_{\alpha}
d^{-}_{\beta}+2Ud^{-}_{\nu}U^{\da}\partial_{\alpha}d^{+}_{\beta}
-2\partial_{\nu}UU^{+}d^{+}_{\alpha}d^{+}_{\beta}-2\partial_{\nu}
U^{\da}Ud^{-}_{\alpha}d^{-}_{\beta}) \nonumber \\
 & &-{1\over g^{2}}(2Ud^{-}_{\nu}U^{\da}d^{+}_{\alpha}d^{+}_{\beta}+
2U^{\da}d^{+}_{\nu}Ud^{-}_{\alpha}d^{-}_{\beta}+d^{-}_{\nu}
d^{-}_{\alpha}d^{-}_{\beta}+d^{+}_{\nu}d^{+}_{\alpha}d^{+}_{\beta})\}.
\end{eqnarray}
{}From this lagrangian, the same conclusion with ref.[14] has been
reached that the decays of $f_{1}
\rightarrow\rho\rho$ and $\omega\omega$ are forbidden. Therefore,
$f_{1}$ meson can not decay to two real photons. This is
Yang's theorem[38].

The lagrangians of the decay of $f\rightarrow a_{1}\pi$ and
$f\rightarrow\rho\pi\pi$ have been found from eq.(110)
\begin{eqnarray}
\lefteqn{{\cal L}_{fa_{1}\pi}=-\frac{4N_{c}}{3(4\pi)^{2}f_{\pi}}
{1\over g^{2}}(1-{1\over 2\pi^{2}g^{2}})^{-1}
\varepsilon^{\mu\nu\alpha\beta}f_{\mu}\partial_{\nu}
\pi^{i}\partial_{\alpha}a^{i}_{\beta}},\nonumber \\
&&{\cal L}_{f\rho\pi\pi}=\frac{4N_{c}}{3(4\pi)^{2}g^{2}f^{2}_{\pi}}
(1-{4c\over g})(1-{1\over 2\pi^{2}g^{2}})^{-{1\over 2}}\varepsilon
^{\mu\nu\alpha\beta}\epsilon_{ijk}\nonumber \\
&&\{\partial_{\alpha}(\partial_{\nu}
\pi_{i}\pi_{j}\rho^{k}_{\beta})-2\partial_{\nu}\pi_{i}\partial_{
\alpha}\pi_{j}\rho^{k}_{\beta}\}.
\end{eqnarray}
In both ${\cal L}_{f\rho\pi\pi}$ and ${\cal L}_{fa_{1}\pi}$ there are
a factor of ${1\over \pi^{2}}$. Therefore, this theory predicts a
narrower width for $f\rightarrow\rho\pi\pi$. On the other hand,
the factor of $1-{4c\over g}$ in ${\cal L}_{f\rho\pi\pi}$
is very small, hence the process $f\rightarrow a_{1}\pi\rightarrow
\rho\pi\pi$ is dominant the decay of $f\rightarrow\rho\pi\pi$.
Using eqs.(61,111) the decay width has been calculated
\begin{equation}
\Gamma(f_{1}\rightarrow\rho\pi\pi)=6.01 MeV.
\end{equation}
The experimental value of the decay width is $6.96(1\pm 0.33)$ MeV.
The prediction of $\Gamma(f_{1}\rightarrow\rho\pi\pi)$ is
in agreement with data.

In another decay of $f_{1}(1285)$ meson,
$f\rightarrow\eta\pi\pi$(excluding $a_{0}(980)\pi$),
even number of
$\gamma_{5}$ are involved. Therefore, the effective lagragian of
this decay should be found from the lagrangian ${\cal L}_{RE}$(13).
Of course, the vertex of $f\eta\pi\pi$ can also be found from
eq.(108) and it should be the same with the one obtained from
eq.(13).
The calculation shows that in the lagrangian ${\cal L}_{RE}$
the term at the second order in derivatives
\[\frac{F^{2}}{16}TrD_{\mu}UD^{\mu}U^{\da}\]
does not contribute to this decay. The effective lagrangian
of this decay comes from the terms at the fourth order in
derivatives which have a factor of ${1\over (4\pi)^{2}}$. Therefore,
this theory predicts a narrow width for the process $f_{1}
\rightarrow
\eta\pi\pi$. In the chiral limit, the effective lagrangian
has been found from eq.(13)
\begin{equation}
{\cal L}_{f\eta\pi\pi}=\frac{4N_{c}}{3(4\pi)^{2}f^{3}_{\pi}}
{1\over g}(1-{1\over 2\pi^{2}g^{2}})^{-{1\over 2}}(1-{2c\over g})^{3}
0.7104f_{\mu}\{\partial^{\mu}\eta\partial_{\nu}\pi_{i}\partial^{\nu}
\pi_{i}+2\partial_{\nu}\eta\partial^{\nu}\pi_{i}\partial^{\mu}\pi_{i}
\},
\end{equation}
where the factor 0.7104 is from the mixing between $\eta$ and $\eta'$
\(0.7104={1\over \sqrt{3}}(cos\theta -\sqrt{2}sin\theta)\) and \(
\theta=-10^{0}\). The numerical result of the decay width is
\begin{equation}
\Gamma(f\rightarrow\eta\pi\pi)=27.5keV.
\end{equation}
Theoretical prediction of the branch ratio is $1.15\times 10^{-3}
(1\pm 0.13)$ and the data of the branch ratio is $(10^{+7}_{-6})\%$.
In principle, the meson $a_{0}(980)(1^{-}(0^{++}))$ can be incorporated
into the lagrangian, then the decay of $f\rightarrow a_{0}\pi$ can be
studied. However, this is beyond the scope of present paper.

Using VMD and ${\cal L}_{f\rho\pi\pi}$, the decay width
of $f\rightarrow\gamma\pi\pi$ has been calculated
\begin{equation}
\Gamma(f\rightarrow\gamma\pi\pi)=18.5 keV.
\end{equation}
In this theory the decay of $f_{1}\rightarrow virtual photon
+\rho$ involves
loop diagrams whose calculation is beyond the scope of this paper.

{\large\bf The decays of $\rho\rightarrow\eta\gamma$ and $\omega
\rightarrow\eta\gamma$}\\
According to VMD, the decays of  $\rho\rightarrow\eta\gamma$ and
$\omega\rightarrow\eta\gamma$ are related to the vertices
$\eta\rho
\rho$ and $\eta\omega\omega$ in which odd number of $\gamma_{5}$ are
involved and these vertices can not be found from
${\cal L}_{RE}$(eq.(13)). From the lagrangian(1) it can be seen that
the interaction between $\eta$ and other mesons can be found from
\begin{eqnarray}
{\cal L}_{\eta}=-{2i\over f_{\pi}}0.7104m\eta<\bar{\psi}
\gamma_{5}\psi>,\nonumber \\
<\bar{\psi}\gamma_{5}\psi>={-i\over (2\pi)^{D}}\int d^{D}p
Tr\gamma_{5}s_{F}(x,p).
\end{eqnarray}
Substituting the solution(42) into the equation(116)
${\cal L}_{\eta}$
has been obtained. The leading terms come from \(n=4\).
We are only interested in the terms containing the vertices $\eta vv$,
which have been found
\begin{eqnarray}
\lefteqn{<\bar{\psi}\gamma_{5}\psi>=\frac{N_{c}}{(4\pi)^{2}}
{i\over 6mg}
\varepsilon^{\mu\nu\alpha\beta}Tr(F^{+}_{\mu\nu}D^{-}_{\alpha}
D^{+}_{\beta}+F^{-}_{\mu\nu}D^{+}_{\alpha}D^{-}_{\beta})}\nonumber \\
 & &+\frac{N_{c}}{(4\pi)^{2}}{i\over 6mg}\varepsilon^{\mu\nu\alpha\beta}
Tr(D^{+}_{\mu}D^{-}_{\nu}F^{+}_{\alpha\beta}+D^{-}_{\mu}D^{+}
_{\nu}F^{-}_{\alpha\beta})\nonumber \\
 & &+\frac{N_{c}}{(4\pi)^{2}}{i\over 3m}\varepsilon^{\mu\nu\alpha\beta}
Tr(D^{+}_{\mu}D^{-}_{\nu}D^{+}_{\alpha}D^{-}_{\beta}
+D^{-}_{\mu}
D^{+}_{\nu}D^{-}_{\alpha}D^{+}_{\beta}), \nonumber \\
& &D^{\pm}_{\mu}=\partial_{\mu}-{i\over g}d^{\pm}_{\mu},\;\;
d^{\pm}_{\mu}=v_{\mu}\mp a_{\mu},\;\;\;a_{\mu}\rightarrow(1-
{1\over 2\pi^{2}g^{2}})^{-{1\over 2}}a_{\mu}-c\partial_{\mu}\pi,
\nonumber \\
& &F^{\pm}_{\mu\nu}=\partial_{\mu}d^{\pm}_{\nu}-\partial_{\nu}d^{\pm}
_{\mu}-{i\over g}[d^{\pm}_{\mu}, d^{\pm}_{\nu}].
\end{eqnarray}
The vertices of $\eta\rho\rho$ and $\eta\omega\omega$ are
revealed from eqs.(117)
\begin{equation}
{\cal L}_{\eta vv}=-{8\over g^{2}f_{\pi}}\frac{N_{c}}{(4\pi)^{2}}
\varepsilon^{\mu\nu\alpha\beta}0.7104\eta(\partial_{\mu}\rho^{i}_{\nu}
\partial_{\alpha}\rho^{i}_{\beta}+\partial_{\mu}\omega_{\nu}
\partial_{\alpha}\omega_{\beta}).
\end{equation}
Using VMD we obtain
\begin{eqnarray}
{\cal L}_{\rho\eta\gamma}=-{8e\over gf_{\pi}}\frac{N_{c}}{(4\pi)^{2}}
\varepsilon^{\mu\nu\alpha\beta}0.7104\eta\partial_{\mu}\rho^{0}_{\nu}
\partial_{\alpha}A_{\beta},\nonumber \\
{\cal L}_{\omega\eta\gamma}=-{8e\over 3gf_{\pi}}\frac{N_{c}}
{(4\pi)^{2}}
\varepsilon^{\mu\nu\alpha\beta}0.7104\eta\partial_{\mu}
\omega_{\nu}
\partial_{\alpha}A_{\beta}.
\end{eqnarray}
The decay widths of $\rho\rightarrow\eta\gamma$ and $\omega
\rightarrow\eta\gamma$ have been calculated
\begin{eqnarray}
\Gamma(\rho\rightarrow\eta\gamma)=46.1keV,\;\;
B(\rho\rightarrow\eta\gamma)
=3.04\times 10^{-4},\nonumber \\
\Gamma(\omega\rightarrow\eta\gamma)=5.87keV,\;\;
B(\omega\rightarrow\eta\gamma)
=6.96\times 10^{-4}.
\end{eqnarray}
The experimental data are
\begin{equation}
B(\rho\rightarrow\eta\gamma)=(3.8\pm 0.7)\times 10^{-4},\;\;
B(\omega\rightarrow\eta\gamma)=(8.3\pm 2.1)\times 10^{-4},\;\;
\end{equation}
Theoretical predictions are in good agreement with data.
For $\eta\rightarrow\gamma\gamma$
besides $\eta\rightarrow\rho\rho$ and $\eta\rightarrow
\omega\omega$, the process $\eta\rightarrow\phi\phi$ also contributes
to $\eta\rightarrow\gamma\gamma$, we will study $\eta\rightarrow
\gamma\gamma$ in another paper in which the strange flavor is
included.

After the study of those physical processes, three problems
of this theory should be discussed. They are: loop diagrams,
dynamical chiral symmetry breaking, and momentum expansion.

{\large\bf Large $N_{c}$ expansion}\\
According to t'Hooft[6], in large $N_{c}$ limit $QCD$ is equivalent
to a meson theory at low energy. Therefore, large $N_{c}$
expansion plays a crucial role in the connection between $QCD$ and
effective meson theory, even though we do not know how to derive
the lagrangian of effective meson theory from $QCD$ directly.
In present theory the large $N_{c}$ expansion plays an important
role too. The quark fields in lagrangian(1) carry
colors. In order to obtain the effective lagrangian of mesons
from the lagrangian(1), the quark fields have been integrated out
by path integral. After this integration the trace in the color
space generates the number of color $N_{c}$.
Parameter m of the lagrangian(1) is $O(1)$ in large $N_{c}$
expansion.
Eq.(14) determines that $F^{2}$ is order of $N_{c}$, hence $f_{\pi}$
is order of $O(\sqrt{N_{c}})$.
The coupling constant g defined by eq.(15) is $O(\sqrt{N_{c}})$.
After normalization, the physical meson fields, pion, $\eta$,
$\rho$, $\omega$, $a_{1}$, and $f_{1}$ are all at order of
$O(\sqrt{N_{c}})$. It is needed to point out that the original form
of the factor $(1-{1\over 2\pi^{2}g^{2}})^{-{1\over 2}}$ in the
normalization of axial-vector meson is $(1-{{N_{c}}\over 6\pi^{2}
g^{2}})^{-{1\over 2}}$. Therefore, this factor is at order of $O(1)$.
The masses of mesons are at order of $O(1)$.
Using all these results, it is not difficult to find out that all
the vertices of this paper are at order of $N_{c}$ and it is obvious
that the propagator of meson is order of $O(1)$. Therefore,
order of magnitude of a Feynman diagram of mesons in large $N_{c}$
expansion is given by
\begin{equation}
N_{c}^{N_{v}-N_{p}},
\end{equation}
where $N_{v}$ is the number of vertices and $N_{p}$ is the number of
internal lines.
Eq.(122) tells that all tree diagrams are at order of $O(N_{c})$, hence
they are leading contributions. A diagram with loops is at higher order
in large $N_{c}$ expansion. For instance, a diagram of one loop with
two internal lines is order of $O(1)$. In this paper all calculations
have been done at tree level. Most of the theoretical predictions
are in agreement with data.  This success can be viewed as a
support of the large $N_{c}$ expansion. However, in some cases due to
cancellation, for instance the direct coupling of $\omega\rightarrow
3\pi$, or other reasons the leading term is very small a correction
of loop diagram should be taken.

{\large\bf Dynamical chiral symmetry breaking}\\
The parameter m in eq.(1) is associated with
quark condensate which is defined as
\begin{equation}
<0|\bar{\psi}(x)\psi(x)|0>=-{i\over (2\pi)^{D}}\int d^{D}p
Tr<0|s_{F}(p,x)|0>.
\end{equation}
At tree level, using eq.(42) we obtain the relation between m and
quark condensate
\begin{equation}
<0|\bar{\psi}(x)\psi(x)|0>=3m^{3}g^{2}(1+{1\over 2\pi^{2}g^{2}}).
\end{equation}
This is $u$ and $d$ quark condensate.
Nonzero quark condensate means dynamical chiral symmetry breaking.
Therefore, there is dynamical chiral symmetry breaking in this theory.
On the other hand, the quark mass term $-\bar{\psi}M\psi$ can be
introduced to the lagrangian(1), where $M$ is the mass matrix of
u and d quark. Using eqs.(40,42) and removing a constant
away the leading term in quark mass expansion has been obtained
\begin{equation}
-<\bar{\psi}M\psi>=\frac{i}{(2\pi)^{D}}\int d^{D}pTrMs_{F}(x,x)=
-{1\over f^{2}_{\pi}}(m_{u}+m_{d})<0|\bar{\psi}
\psi|0>.
\end{equation}
The pion mass(79) revealed from this equation is
\begin{equation}
m^{2}_{\pi}=-{2\over f^{2}_{\pi}}(m_{u}+m_{d})<0|\bar{\psi}\psi|0>.
\end{equation}
Detailed discussion of masses of pseudoscalar mesons can be found
in ref.[31]. Eq.(126)
is a well known formula obtained by the theory of chiral symmetry
breaking proposed by Gell-Mann, Oakes, and Renner[39],
and by Glashow and
Weinberg[40].Eq.(126) tells that the quark condensate is negative,
hence the parameter m is negative too.
The parameter m has been determined from eqs.(15,25,26)
\begin{equation}
m=-300 MeV,
\end{equation}
and eq.(124) determines
\begin{equation}
<0|\bar{\psi}\psi|0>=-(241MeV)^{3}.
\end{equation}
The quark mass is determined to be
\[m_{u}+m_{d}=23.5MeV.\]
It is about twice the current value[41]. As pointed in ref.[41], the
determination of absolute value of quark mass is model dependent.

{\large\bf Derivative expansion}\\
This theory is an effective meson theory at low energies. Like the
chiral perturbation theory[1], the derivative expansion
(to be accurate, covariant derivative expansion) has been applied.
It seems that the derivative expansion works in the studies
presented in this paper. The calculations of decay widths are
good examples. If the terms at the second
order in derivatives in lagrangian(13) contribute to the decay of a
meson, the decay width of this meson is broader. $\rho$ decay and
$a_{1}$ decay are two examples. If only the terms at the fourth order
in derivatives contribute to the decay, the decay width is narrower.
The reason is that in the decay amplitude there is a factor of
${1\over \pi^{2}}$. The predictions of narrower widths for
decays, $\omega\rightarrow 3\pi$,
$f\rightarrow\rho\pi\pi$, and $f\rightarrow\eta\pi\pi$, support
this argument. From table I it can be seen that the scattering lengths
and slopes($a^{0}_{0}$, $b^{0}_{0}$, $a^{1}_{1}$, $a^{2}_{0}$,
$b^{2}_{0}$),
which are obtained from the terms at the zeroth order or the
second order in derivatives, are much greater than $a^{0}_{2}$,
$a^{2}_{2}$, and $b^{1}_{1}$ which are from the terms at the
fourth order in derivatives.

This theory is an effective theory and it is not renormalizable,
as mentioned above, a cut-off of momentum has to be introduced
into this theory. The eq.(15) can be used to determine the cut-off.
Using cut-off instead dimension regularization, eq.(15) has been
rewritten as
\begin{equation}
\frac{N_{c}}{(4\pi)^{2}}\{log(1+{\Lambda^{2}\over m^{2}})+
\frac{1}{1+\frac{\Lambda^{2}}{m^{2}}}-1\}={1\over 16}\frac{F^{2}}
{m^{2}}={3\over 8}g^{2}.
\end{equation}
Using the values of m and g, we obtain
\begin{equation}
\Lambda=1.6GeV.
\end{equation}
Derivative expansion is momentum expansion and $\Lambda$ is the
maximum momentum. The momentum expansion requires that the momentum
is less than $\Lambda$. The masses of $\rho$, $\omega$,
$a_{1}$, and $f_{1}$ are less than $\Lambda$.
However, there is physical case in which momentum expansion
is not suitable. In $\pi\pi$ scattering there is $\rho$ resonance in
the scattering amplitudes(84).
At very low energy the momentum expansion has been applied
to $\pi\pi$ scattering amplitudes, however, in the region of $\rho$
resonance the momentum expansion is not working because it
destroys the resonance. Therefore, we do not apply the momentum
expansion to the amplitudes of resonance in this paper.

{\large\bf Summary of the results}\\
To summarize the results achieved by this theory. VMD has been revealed
from this theory. Weinberg's first sum rule and new relations about
$a_{1}$ meson decay are satisfied. KSFR sum rule is satisfied
pretty well. New mass relations between vector and axial-vector mesons
have been found. Weinberg's $\pi\pi$ scattering lengths and slopes
have been revealed. The amplitude of $\pi^{0}\rightarrow\gamma
\gamma$ obtained by this theory is exact the same with the prediction
of triangle anomaly. This theory provides a united description of
the physical processes with normal and abnormal parity and the
universality of coupling has been revealed.
The effective lagrangians used to treat the processes of abnormal
parity are exactly the same with the ones obtained
from gauging Wess-Zumino lagrangian with vector and axial-vector
mesons. In chiral limit, we take $f_{\pi}$ and $m_{\rho}$ as inputs.
The coupling constant g is chosen to fit the data. In
phase spaces physical $m_{\pi}$ and $m_{\eta}$($m_{\pi}$ and
 $m_{\eta}$ are inputs) have been taken and in the amplitudes
only the leading terms have been kept in chiral perturbation.
The results of $\pi\pi$ scattering have been shown in Table I and
Fig.1. Other results are listed in Table II.
\begin{table}[h]
\begin{center}
\caption {Table II Summary of the results}
\begin{tabular}{|c|c|c|} \hline
    &  Experimental  &  Theoretical  \\ \hline
$f_{\pi}$   & 0.186GeV    & input         \\ \hline
$m_{\rho}$  & 769.9$\pm$ 0.8MeV     & input         \\ \hline
$m_{\pi}$   & 0.138 Gev    & input         \\ \hline
$m_{\eta}$  & 547.45$\pm$0.19MeV    & input         \\ \hline
g         &              & 0.35  input   \\ \hline
$m_{\omega}$& 781.94$\pm$0.12MeV    & 0.77 GeV     \\ \hline
$m_{a}$     & 1230$\pm$40MeV    & 1.389 GeV     \\ \hline
$m_{f_{1}}$     & $1282\pm5$MeV    & 1.389 GeV     \\ \hline
$\pi$ form factor & consistent with $\rho$ pole & $\rho$ pole
\\ \hline
radius of $\pi$ & $0.663\pm$0.023fm & 0.63 fm        \\ \hline
$g_{\rho\gamma}$ & $0.116(1\pm 0.05)$ $GeV^{2}$
& 0.104 $GeV^{2}$ \\ \hline
$g_{\omega\gamma}$ & $0.0359(1\pm 0.03)$ $GeV^{2}$ & 0.0357 $GeV^{2}$
\\ \hline
$\Gamma(\rho\rightarrow\pi\pi)$ & $151.2\pm 1.2$ MeV & 135. MeV \\
\hline
$\Gamma(\omega\rightarrow\pi\pi)$ & $0.186(1\pm 0.15)$MeV & 0.136MeV
\\  \hline
\end{tabular}
\end{center}
\end{table}
\begin{table}
\begin{center}
\begin{tabular}{|c|c|c|} \hline
$\Gamma(a_{1}\rightarrow\rho\pi)$ & $\sim$400 MeV & 325 MeV \\
\hline
$\Gamma(a_{1}\rightarrow\gamma\pi)$ &(640$\pm$246)keV & 252keV \\
\hline
${d\over s}(a_{1}\rightarrow\rho\pi)$ & $-0.11\pm 0.02$ &-0.097 \\
\hline
$\Gamma(\tau\rightarrow a_{1}\nu)$ &$(2.42\pm 0.76)\times 10^{-13}$
GeV    &$1.56\times 10^{-13}$GeV   \\ \hline
$\Gamma(\tau\rightarrow \rho\nu)$ &$(0.495\pm 0.023)\times
10^{-12}$GeV   & 4.84$\times 10^{-13}$GeV \\  \hline
$\Gamma(\pi^{0}\rightarrow\gamma\gamma)$ & $7.74(1\pm 0.072) $eV
&7.64eV \\ \hline
a(form factor of $\pi^{0}\rightarrow\gamma\gamma$) & 0.032$\pm$
0.004 &0.03 \\ \hline
$\Gamma(\omega\rightarrow\pi\gamma)$ & 717(1$\pm$0.07)keV
& 724 keV \\ \hline
$\Gamma(\rho\rightarrow\pi\gamma)$ & 68.2(1$\pm$0.12)keV &
76.2keV \\  \hline
$\Gamma(\omega\rightarrow\pi\pi\pi)$ & 7.43(1$\pm$
0.02)MeV  &5 MeV \\ \hline
$\Gamma(f_{1}\rightarrow\rho\pi\pi)$ &
6.96(1$\pm$0.33)MeV&6.01MeV\\ \hline
$B(f_{1}\rightarrow\eta\pi\pi)$ &$(10^{+7}_{-6})\%$
&1.15$\times 10^{-3}$ \\ \hline
$\Gamma(f_{1}\rightarrow\gamma\pi\pi)$ &        &18.5keV  \\ \hline
$B(\rho\rightarrow\gamma\eta)$ &$(3.8\pm 0.7)\times 10^{-4}$
&3.04$\times 10^{-4}$\\ \hline
$B(\omega\rightarrow\gamma\eta)$ &$(8.3\pm 2.1)
\times 10^{-4}$ &$6.96\times 10^{-4}$ \\ \hline
\end{tabular}
\end{center}
\end{table}

The author likes to thank G.P.Lepage for helphul discussion and to
thank
T.Barnes and E.Swanson for help.
This research is partially
supported by DE-91ER75661.

\end{document}